\documentclass[11pt]{JHEP3}

\usepackage{mathpazo}
\usepackage{amsmath}
\usepackage{amssymb}
\usepackage{graphicx}
\usepackage{epstopdf}
\usepackage{cite}

\newcommand{\be}{\begin{eqnarray}}
\newcommand{\ee}{\end{eqnarray}}
\newcommand{\nn}{\nonumber}
\newcommand{\bn}{\begin{enumerate}}
\newcommand{\en}{\end{enumerate}}

\def\identity{{\rlap{1} \hskip 1.6pt \hbox{1}}}

\def\Ds{D \hskip -7pt / \hskip 2pt}

\def\CB{{\cal B}}

\def\CF{{\cal F}}

\def\CL{{\cal L}}

\def\CN{{\cal N}}
\def\CO{{\cal O}}
\def\CP{{\cal P}}
\def\CQ{{\cal Q}}

\def\a{\alpha}
\def\b{\beta}
\def\g{\gamma}
\def\d{\delta}
\def\e{\epsilon}
\def\ve{\varepsilon}

\def\th{\theta}

\def\l{\lambda}
\def\m{\mu}
\def\n{\nu}

\def\r{\rho}

\def\s{\sigma}

\def\t{\tau}

\def\w{\omega}
\def\G{\Gamma}

\def\L{\Lambda}
\def\S{\Sigma}

\def\O{\Omega}

\def\dd{\rm d}

\def\half{\frac{1}{2}}
\def\thalf{{\textstyle \frac{1}{2}}}

\def\goto{\rightarrow}

\def\p{\partial}

\def\identity{{\rlap{1} \hskip 1.6pt \hbox{1}}}

\def\tr{{\rm tr}}

\def\da{{\dot{\a}}}
\def\db{{\dot{\b}}}
\def\dg{{\dot{\g}}}
\def\dd{{\dot{\d}}}
\def\tq{{\tilde{q}}}
\def\tps{{\tilde{\psi}}}
\def\jmath{{j}}

\newcommand{\sla}[1]{{/\!\!\!\!{#1}}}

\title{Topological Chern-Simons Sigma Model}

\author{Eunkyung Koh$^1$,  Sangmin Lee$^1$, Sungjay Lee$^2$
\\
\\
$^1$Department of Physics and Astronomy, Seoul National University,
Seoul 151-747, Korea
\\
$^2$School of Physics, Korea Institute for Advanced Study, Seoul 130-012, Korea
\\
\\
E-mail:
\email{ekoh@phya.snu.ac.kr, sangmin@snu.ac.kr, sjlee@kias.re.kr}
}

\abstract{
We consider topological twisting of
recently constructed Chern-Simons-matter theories in three dimensions with $\CN = 4$ or higher supersymmetry.
We enumerate physically inequivalent twistings
for each $\CN$, and find two different twistings for $\CN=4$,
one for $\CN=5,6$, and four for $\CN=8$.
We construct the two types of $\CN=4$ topological theories, which we call A/B-models, in full detail.
The A-model has been recently studied by Kapustin and Saulina.
The B-model is new and it consists solely of a Chern-Simons term
of a complex gauge field up to BRST-exact terms.
We also compare the new theories
with topological Yang-Mills theories
and find some interesting connections.
In particular, the A-model seems to offer a new perspective on Casson invariant
and its relation to Rozansky-Witten theory.
}


\preprint{KIAS-P09037}

\begin{document}


\section{Introduction}

For the past two years, a large class of new Chern-Simons-matter theories
has been discovered. Since the seminal work of
Bagger and Lambert \cite{BL2} and Gustavsson\cite{gus1} (BLG),
where a theory with maximal $\CN=8$ was first constructed,
the list of supersymmetric Chern-Simons theories has expanded quickly.

Gaiotto and Witten \cite{gw} (GW) gave a general prescription for
coupling Chern-Simons theory to hyper-multiplets, allowing for
construction of a large class of new theories at once.
It was soon augmented \cite{n4} by adding twisted hyper-multiplets, so that
all $\CN \ge 4$ theories can be understood in a unified way in the GW framework.
\footnote{
We will focus exclusively on $\CN \ge 4$ theories. 
See, for instance, some early works  \cite{gates1,gates2,kao} and a recent review \cite{gy} for $\CN \le 3$ theories,}

The scalar fields of $\CN=4$ theories can describe a sigma model with a hyper-K\"ahler target space.
The $\CN\ge 5$ theories have much less freedom for their target spaces: flat spaces and their orbifolds. An exhaustive list of $\CN \ge 5$ theories can be found in \cite{n56,BL6,tachi6}.

The main goal of this paper is to study topological field theories
obtained by twisting the new Chern-Simons-matter theories.
There are two well-known topological theories in three dimensions:
pure Chern-Simons theory \cite{purecs} and Rozansky-Witten theory \cite{rw}.
The latter is a twisted version of $\CN=4$ sigma model without gauge symmetry.
Clearly, topological twisting of Chern-Simons-matter theories would lead to a mixture
of pure Chern-Simons and Rozansky-Witten theories.
It is conceivable that the mixed theory may shed light on
relations between the two seemingly different theories.
In this paper, we take a first step toward understanding the new topological theories.

In section 2, we begin by writing down the physical $\CN=4$ theories
in full generality. It was already done in \cite{n56} for flat target space.
For non-linear hyper-K\"ahler target spaces, the recipe was given
in \cite{gw}, but the explicit form of the Lagrangian and supersymmetry transformation rules were not available in the literature.

We then proceed to topological twisting in section 3.
First, 
we enumerate all possible inequivalent twistings for $\CN \ge 4$ theories,
\footnote{
Note that $\CN=4$ is the minimal number of supersymmetry for twisting.
The $\CN=3$ case is excluded because the supercharges form
a triplet under the $SO(3)$ $R$-symmetry and a doublet under the Lorentz symmetry, so that none of the supercharges become scalar under the twisted Lorentz symmetry.
}
summarized in the following table:

\vskip 3mm

\begin{table}[htbp]
\begin{center}
\begin{tabular}{c|c}
$\CN=4$ (hyper-multiplet only) & A/B \\
$\CN=4$ (hyper + twisted hyper), $5, 6$ & AB \\
$\CN=8$ & AB/C/D/E \
\end{tabular}
\end{center}
\caption{Topological twistings of $\CN\ge 4$ Chern-Simons-matter theories.}
\label{twi-all}
\end{table}

\noindent
For $\CN=4$ theories with $SO(4)\simeq SU(2)_L \times SU(2)_R$ $R$-symmetry,
we can use either of the two $SU(2)$ factors for twisting
with the Lorentz group $SU(2)_E$.
When the $\CN=4$ theory contains hyper-multiplets but no twisted hyper-multiplets, the two choices lead to different topological theories,
which we call A and B models.
The distinction disappears when the theory contains both types of hyper-multiplets, which is automatically true of $\CN=5,6$ theories.
We call the result AB-model. For $\CN=8$ theories,
the triality of $SO(8)$ offers three new possibilities aside from the AB-model,
which we call C/D/E models.
The C-twisting of the BLG theory has been considered in \cite{Lee:2008cr},
while the A-model has been studied in a recent paper \cite{ks}
which have some overlap with the current work.

In the remainder of section 3, we explicitly construct the A/B/AB-models
in the $\CN=4$ notation and study their properties. The scalar super-charges
of the A-model are nilpotent up to a gauge transformation, while
nilpotency in the B-model holds up to the equations of motion
for the fermions. A BRST gauge fixing and introduction of auxiliary fields
renders the super-charges fully nilpotent. Topological invariance
of the resulting Lagrangian is verified in the usual manner
by splitting the Lagrangian into a manifestly metric-independent part
and a $Q$-exact part.

In section 4, we take a preliminary step toward 
the computation of topological invariants from the new theories. 
Inspired by the Mukhi-Papageorgakis (MP) map \cite{mukhi} relating
Chern-Simons and Yang-Mills theories, we compare our 
new theories with more well-known topological theories in the literature, 
and argue that the linear A-model with gauge group $SU(2)\times SU(2)$ 
should capture the Casson invariant. The MP map also suggests that
A-model is naturally related to the Rozansky-Witten theory \cite{rw} as well.
We also speculate briefly on how the A and B-models may make contact with the pure Chern-Simons theory \cite{purecs}.

\section{Chern-Simons sigma model}











In this section, we write down
the most general form of Chern-Simons sigma model in $(2+1)$ dimensions.
We first review the linear model constructed in \cite{gw} and extended in \cite{n4}. Then we write down the non-linear model following
the prescription given in \cite{gw}.
We will mostly follow notations of \cite{n4}, except for
an overall rescaling of matter fields.

\subsection{Linear model}

We start with an $Sp(2n)$ group and let $A,B$ indices run over a $2n$-dimensional representation.
We denote the anti-symmetric invariant tensor of $Sp(2n)$ by $\omega_{AB}$ and choose all the generators $t^A_{~B}$ to be anti-Hermitian $(2n\times 2n)$ matrices, such that $t_{AB} \equiv \omega_{AC}t^C_{~B}$ are symmetric matrices. We consider a  Chern-Simons gauge theory whose gauge group is a subgroup of $Sp(2n)$ and we denote the anti-Hermitian generators of the gauge group by $(t^m)^A_{~B}$ which satisfy the commutation relations,
\be
 [t^m,t^n]=f^{mn}{}_p t^p \,.
\ee
Gauge fields are denoted by $(A_m)_\mu$ and
the adjoint indices are raised or lowered by an invariant quadratic
form $k^{mn}$ or its inverse $k_{mn}$ of the gauge group.

We couple the gauge theory with a
hyper-multiplet matter fields  $(q^A_\a, \psi^A_\da)$ satisfying the reality condition
\be
 (q^A_\a)^* =\epsilon^{\a\b}\omega_{AB}q^B_\b,~~~~~~
 (\psi^A_\da)^*=\epsilon^{\da\db}\omega_{AB}\psi^B_\db.
\ee
We use $(\a,\b ; \da,\db)$ doublet indices for the $SU(2)_L\times SU(2)_R$ $R$-symmetry group.

The necessary and sufficient condition for $\CN=4$
supersymmetry \cite{gw} is that $t^m_{AB}$ satisfy
the ``fundamental identity'',
\be
\label{fundi}
k_{mn} t^m_{(AB}t^n_{C)D}=0 ,
\ee
where the indices $A,B,C$ are symmetrized over cyclic permutations.
This identity can be
understood \cite{gw} as the Jacobi identity for three fermionic generators
of a Lie super-algebra,
\begin{equation}
\label{lsa}
  [M^m,M^n]=f^{mn}_{~~~p}M^p,~~~~
  [M^m,Q_A]=Q_B (t^m)^B_{~A},~~~~
  \{Q_A,Q_B\}= t^m_{AB} M_m.
\end{equation}
This turns out to be a rather strong constraint on the field content
of the theory.
Namely, the gauge group and the matter content should be such that the gauge
symmetry algebra can be extended to a Lie super-algebra by adding
fermionic generators in one-to-one correspondence with hyper-multiplets.

To write down the Lagrangian in a manifestly $
\CN=4$ covariant form, it is useful to introduce
the ``moment map" multiplet,
\be
\mu^m_{\a\b}\equiv t^m_{AB}q^A_\a q^B_\b\,,
\;\;\;\;\;
j^m_{\a\db}\equiv t^m_{AB}q^A_\a \psi^B_\db \,,
\;\;\;\;\;
\rho^m_{\da\db}\equiv t^m_{AB}\psi^A_\da \psi^B_\db \,.
\label{m-map}
\ee
As for the Chern-Simons term in the Lagrangian, we use the notation
\be
\CL_{\rm CS} (A) \equiv \varepsilon^{\m\n\l}
 \left(k_{mn} A^m_\m \partial_\n A^n_\l
      +\frac13 f_{mnp} A^m_\m A^n_\n A^p_\l \right)
\ee
As our discussion in this paper will be mostly classical,
we will suppress an overall coefficient of the Lagrangian,
which should satisfy an integrality condition to make
the quantum theory well-defined.

Collecting all notations, we can summarize
the Lagrangian of the Gaiotto-Witten model,
\begin{eqnarray}
 \CL &=& \CL_{\rm CS}(A) + \omega_{AB}
  \left(-\epsilon^{\a\b} D q^A_\a D q^B_\b+i\epsilon^{\da\db}\psi^A_\da \Ds\psi^B_\db \right)
 \nn\\&&
 -i k_{mn} \e^{\a\b} \e^{\dg\dd} \jmath^m_{\a\dg} \jmath^n_{\b\dd}  - \frac{1}{12} f_{mnp} (\mu^m)^\a_{~\b}(\mu^n)^\b_{~\g}(\mu^p)^\g_{~\a}\,,
\label{Lful1}
\end{eqnarray}
and its supersymmetry transformation rules
\be
\label{susytr1}
 &&\delta q_\a^A = i\eta_\a{}^{\da} \psi_\da^A\,,~~~~
 \delta A^m_\m =  i \eta^{\a\da} \gamma_\m \jmath^m_{\a\da}\,,
  \nn \\
 &&\delta\psi_\da^A=
 \left[\Ds q_\a^A + \frac{1}{3}k_{mn}(t^m)^A_{~B} q^B_\b (\mu^n)^\b_{~\a}\right]
 \eta^\a_{~\;\da}\,.
\ee
The supersymmetry parameter $\eta$ transforms in the $(\mathbf{2,2})$ representation of $SU(2)_L \times SU(2)_R$
and satisfies the reality condition\footnote{
In \cite{n56}, the same reality condition was stated with a wrong sign.}
\be
\label{eta-real1}
(\eta_\a{}^\da)^*
= - \e^{\a\b} \e_{\da\db} \eta_\b{}^\db \,.
\ee

To obtain the most general $\CN=4$ Chern-Simons (linear) sigma model,
one should add twisted hyper-multiplets $(\tilde{q}^A_\da, \tilde{\psi}^A_\a)$ to the Gaiotto-Witten model \cite{n4}.
The gauge generators $\tilde{t}^m_{AB}$ also satisfy
the fundamental identity (\ref{fundi})
and define the twisted moment map multiplet
similar to (\ref{m-map}).
It is also useful to introduce yet another notation,
\be
\mu^{mn} = \e^{\a\b} (t^mt^n)_{AB} q^A_\a q^B_\b \,,
\;\;\;
\tilde{\mu}^{mn} = \e^{\da\db} (\tilde{t}^m \tilde{t}^n)_{AB} \tq^A_\da \tq^B_\db \,.
\label{mumu1}
\ee
The full Lagrangian is given by
\begin{eqnarray}
 \CL &=&
\CL_{\rm CS}(A)
 +\omega_{AB}
  \left(-\epsilon^{\a\b} D q^A_\a D q^B_\b
        +i\epsilon^{\da\db}\psi^A_\da \Ds\psi^B_\db \right)
 + \omega_{AB}
  \left(-\epsilon^{\da\db} D \tilde q^A_\da D \tilde q^B_\db
       +i\epsilon^{\a\b}\tilde\psi^A_\a \Ds\tilde\psi^B_\b \right)
 \nn\\&&
 -i k_{mn} \left( \e^{\a\b} \e^{\dg\dd}\jmath^m_{\a\dg}\jmath^n_{\b\dd}
 + \e^{\da\db} \e^{\g\d}\tilde\jmath^m_{\da\g}\tilde\jmath^n_{\db\d}
 +4  \epsilon^{\a\g}\epsilon^{\db\dd}
   \jmath^m_{\a\db}\tilde\jmath^n_{\dd\g}
   -\epsilon^{\da\dg}\epsilon^{\db\dd}
 \tilde\mu^m_{\da\db} \rho^n_{\dg\dd}
 -\epsilon^{\a\g}\epsilon^{\b\d}
  \mu^m_{\a\b}\tilde{\rho}_{\g\d}  \right)
 \nn\\&&
  -\frac{1}{12} f_{mnp} \left( (\mu^m)^\a_{~\b}(\mu^n)^\b_{~\g}(\mu^p)^\g_{~\a}
   +(\tilde\mu^m)^{\dot{\a}}_{~\dot{\b}}
   (\tilde\mu^n)^{\dot{\b}}_{~\dot{\g}}
   (\tilde\mu^p)^{\dot{\g}}_{~\dot{\a}} \right)
 \nn\\&&
  -\half  \tilde\mu^{mn} (\mu_m)^\a_{~\b}(\mu_n)^\b_{~\a}
  -\half  \mu^{mn}  (\tilde\mu_m)^\da_{~\db}(\tilde\mu_n)^\db_{~\da}\,.
\label{Lful4}
\end{eqnarray}
The supersymmetry transformation rules read
\begin{eqnarray}
 && \delta q_\alpha^A
    = i\eta_\alpha^{~\;\dot\alpha}\psi_{\dot\alpha}^A\,,
~~~
    \delta\tilde q_{\dot\alpha}^A
    = i\eta_{~\;\dot\alpha}^{ \alpha}\tilde\psi_\alpha^A\,,
~~~
    \delta A^m_\mu =  i\eta^{\alpha\dot\alpha}\gamma_\mu
            (\jmath^m_{\alpha\dot\alpha}+\tilde\jmath^m_{\dot\alpha\alpha})\,,
 \nn\\
 && \delta\psi_{\dot\alpha}^A
    = \left[\sla Dq_\alpha^A
           +\frac{1}{3}(t_m)^A_{~B}q^B_\beta(\mu^m)^\beta_{~\alpha}\right]
            \eta^\alpha_{~\;\dot\alpha}
           - (t_m)^A_{~B}q^B_\beta(\tilde\mu^m)^{\dot\beta}_{~\dot\alpha}
            \eta^\beta_{~\;\dot\beta} \,,
 \nn\\
 && \delta\tilde\psi_\alpha^A
    = \left[\sla D\tilde q_{\dot\alpha}^A
            +\frac{1}{3}(\tilde t_m)^A_{~B}\tilde q^B_{\dot\beta}
             (\tilde\mu^m)^{\dot\beta}_{~\dot\alpha}\right]
            \eta_\alpha^{~\;\dot\alpha}
           - (\tilde t_m)^A_{~B}\tilde q^B_{\dot\beta}
             (\mu^m)^{\beta}_{~\alpha}
            \eta_\beta^{~\;\dot\beta} \,.
            \label{Sful4}
\end{eqnarray}

\paragraph{Mass deformation}

The $\CN=4$ superconformal Chern-Simons theories allow
a mass-deformation which preserves all of the Poincar\'e supersymmetry
\cite{n4} (See also \cite{n56,gomis,n8}).
For the Gaiotto-Witten model, the deformation amounts to adding
the mass terms and a quartic interaction term to the Lagrangian,
\begin{eqnarray}
\label{d-lag}
  \CL_\text{mass} = - \omega_{AB} k_{mn}\left( m^2
  \e^{\a\b} q_\a^A q_\b^B + i m \e^{\da\db}
  \psi_\da^A \psi_\db^B \right)
  - \frac{2}{3}m\,k_{mn}
  (\mu^m)_{\a\b} (\mu^n)^{\b\a} \ .
\end{eqnarray}
%
One show that the mass-deformed Lagrangian still
preserves the $\CN=4$ supersymmetry, provided that the
supersymmetry transformation rule for the fermion is also modified by
an additional term,
\begin{eqnarray}
  \d_\text{mass} \psi_\da^A = m q_\a^A \eta^\a_{\ \da}\ .
\end{eqnarray}
The deformed supersymmetry algebra contains a non-central
extension,
\begin{eqnarray}
\label{dsusy}
  \big\{ Q^{\a\da} , Q^{\b\db} \big\} = \big( \g^\mu \e^{-1} \big) P_\mu \e^{\a\b}\e^{\da \db}
  + \e^{-1} 2 m  \big( \e^{\a\b} R^{\da\db} - \e^{\da \db} R^{\a\b} \big)\ ,
\end{eqnarray}
where $R^{\a\b}$, $R^{\da\db}$ denote the generators of $SU(2)_L\times SU(2)_R$.

It follows from (\ref{dsusy}) that for the mass deformation
of the general theory with both types of hyper-multiplets,
the mass parameters of the hyper- and twisted hyper-multiplets should be
equal.
The mass-deformed term in the Lagrangian of the general theory turns out to be
the sum of the contributions from
the two types of multiplets.
\begin{eqnarray}
  \CL_\text{mass} &=&
  - \omega_{AB} \Big( m^2
  \e^{\a\b} q_a^A q_b^B  + m^2 \e^{\da\db} \tilde q_\da^A \tilde q_\db^B
  + i m \e^{\da\db}   \psi_\da^A \psi_\db^B - i m \e^{\a\b}
  \tilde \psi_\a^A \tilde \psi_\b^B \Big) \nonumber \\ && \hspace*{0.5cm}
   - \frac{2}{3}m\, k_{mn} \Big( (\mu^m)_{\a\b} (\mu^n)^{\b\a}
   - (\tilde \mu^m)_{\da\db} (\tilde \mu^n)^{\db\da} \Big) \ .
\end{eqnarray}
The supersymmetry transformation rules for fermion fields are again
modified as
\begin{eqnarray}
  \d_\text{mass} \psi_\da^A = m q_\a^A \eta^\a_{\ \da}\ ,
  \qquad
  \d_\text{mass} \tilde \psi_\a^A = m \tilde q_\da^A \eta_\a^{\ \da}\ .
\end{eqnarray}
%


\subsection{Non-linear model} \label{non-linear}

The linear sigma model explained above can be generalized to a non-linear model \cite{gw} whose target space is a hyper-K\"ahler manifold $X$.
The scalar fields are now local coordinates $q^i$ on $X$ ($i=1,\cdots,4n={\rm dim}X$).
We begin this subsection with a brief review of hyper-K\"ahler geometry, closely following \cite{rw},
and move on to describe the sigma model
as explained in \cite{gw}.

\paragraph{Hyper-K\"ahler geometry}
The hyper-K\"ahler structure of $X$ can be described by
the existence of anti-symmetric inner products $\w_{AB}$ and
$\e_{\a\b}$, which leads to three symplectic forms
\be
\O_{\a\b} = \w_{AB} \e_{\a\g} \e_{\b\d} \,e^{A\g} \wedge e^{B\d} \,.
\ee
We introduced the hyper-K\"ahler vielbein satisfying
\be
e^{A\a} = e^{A\a}_i dq^i,
\;\;\;
d e^A + \G^A{}_B \wedge e^B = 0 \,,
\ee
where $\G^{A}{}_B = \G_i{}^A{}_B dq^i$ is the metric connection
in the $Sp(2n)$ holonomy group.
The curvature tensor is given by
\be
R^A{}_B = d\G^A{}_B + \G^A{}_C \wedge \G^C{}_B = \thalf R^A{}_{Bij} dq^i \wedge dq^j \,.
\ee
The hyper-K\"ahler structure and the identity $R_{i[jkl]}=0$ further implies that
\be
R_{AB} = \thalf \O_{ABCD} \e_{\g\d} e^{C\g} \wedge e^{D\d} \,,
\ee
where $\O_{ABCD}$ is totally symmetric.

\paragraph{Target space isometry}
Consider a set of Killing vectors $\{ V^m \}$ on $X$ satisfying the Lie(-bracket) algebra,
\be
\label{Lie-V}
\left[ V^m, V^n \right] = f^{mn}{}_p V^p \,.
\ee
On a K\"ahler manifold, a Killing vector preserving
the complex structure satisfies $\nabla_I V_J = 0$ and
$\nabla_I V_{\bar{J}} = - \nabla_{\bar{J}} V_I$,
where $(I,J;\bar{I},\bar{J})$ are holomorphic and anti-holomorphic indices.
The corresponding statement in the hyper-K\"ahler case is that
$\{V^m\}$ preserving all three complex structures should satisfy
\be
\label{Killing-V}
\nabla_{A\a} V^m_{B\b} = t^m_{AB} \e_{\a\b},
\ee
for some symmetric tensor fields $t^m_{AB}$.

On any Riemannian manifold, the Killing equation and $R_{i[jkl]}=0$ imply
\be
\nabla_k \left( \nabla_i V_j \right) = V^l R_{lkij} \,.
\ee
The hyper-K\"ahler version of the identity can be written as
\be
\nabla_i (t^m)^A{}_B = - R_{ij}{}^A{}_B (V^m)^j.
\label{t-r-iden}
\ee
Differentiating (\ref{Lie-V}) and using (\ref{t-r-iden}), we find
\be
[t^m,t^n]^A{}_B = f^{mn}{}_p (t^p)^A{}_B + R_{ij}{}^A{}_B (V^m)^i (V^n)^j \,.
\label{t-comm}
\ee

The moment maps are defined by
\be
\label{moment-map}
d \left(\m^m_{\a\b}\right) = i_{V^m}(\O_{\a\b}) \,.
\ee
In general, there are undetermined additive constants in $\m$,
which corresponds to the possibility of adding Fayet-Iliopoulos $D$-terms
for $U(1)$ gauge fields.
It is useful to note that one can integrate (\ref{Lie-V}) and use (\ref{moment-map}) to obtain
\be
\label{mom-int}
- \w_{AB} (V^m)^A_{(\a}(V^n)^B_{\b)} = i_{V^m} d\m^{n}_{\a\b}  = f^{mn}{}_p \mu^p_{\a\b} \,.
\ee
The other two components of the moment map multiplet can be defined by
\be
j^m_{\a\db} \equiv - V^m_{A\a} \psi^A_\db \,,
\;\;\;\;\;
\rho^m_{\da\db} \equiv t^m_{AB} \psi^A_\da \psi^B_\db \,.
\ee
In terms of $\m^m_{\a\b}$, the fundamental identity for the non-linear model can be written as
\be
\label{fundi2}
k_{mn} \m^m_{(\a\b} \m^n_{\g\d)} = 0 \,.
\ee
In the non-conformal cases, a weaker condition, which is
a second descendant of the fundamental identity, is sufficient to ensure $\CN=4$ supersymmetry,
\be
\label{fundi3}
k_{mn} \left( \m^m_{\a\b} \rho^n_{\da\db} + j^m_{\a\da} j^n_{\b\db}
+ j^m_{\a\db} j^n_{\b\da} \right) = 0\,.
\ee

\paragraph{Gauging the isometry}
The target space isometry can be gauged by
imposing the following transformation rules on the fields:
\be
\label{dl-q}
\d_\L q^i &=& \L_m (V^m)^i \,,
\\
\label{dl-p}
\d_\L \psi^A &=& -\L_m (\hat{t}^m)^A{}_B \psi^B
\;\equiv\; - \L_m \left[ (t^m)^A{}_B +(V^m)^i\G_i{}^A{}_B \right] \psi^{B} \,,
\\
\d_\L (A_m)_\m &=& D_\m \L_m = \partial_\m \L_m + f^{np}{}_m
(A_n)_\m \L_p  \,.
\ee
The covariant derivatives for the matter fields are defined by
\be
D_\mu q^i &=& \partial_\m q^i - (A_m)_\m (V^m)^i \,,
\\
D_\mu \psi^A &=& \partial_\m \psi^A + \partial_\m q^i \G_i{}^A{}_B \psi^B + (A_m)_\m (t^m)^A{}_B \psi^B
\nn \\
&=& \partial_\m \psi^A + D_\m q^i \G_i{}^A{}_B \psi^B + (A_m)_\m (\hat{t}^m)^A{}_B \psi^B \,.
\ee
They transform homogeneously under the gauge symmetry,
\be
\d_\L\left( D_\mu q^i \right) = \L_n \partial_j(V^n)^i  D_\m q^j \,,
\qquad
\d_\L \left( D_\mu \psi^A \right) =
-\L_m (\hat{t}^m)^A{}_B \psi^B
\,.
\label{d-cov-psi}
\ee
The moment map multiplet also transforms as expected,
\be
\label{mom-tr}
\d_\L \left( \m^m_{\a\b} , j^m_{\a\da}, \rho^m_{\da\db} \right)
= - f^{mn}{}_p \L_n \left( \m^p_{\a\b},j^p_{\a\da}, \rho^p_{\da\db} \right) \,.
\ee
To verify (\ref{d-cov-psi}) and (\ref{mom-tr}),
one needs to use the identities (\ref{t-r-iden}), (\ref{t-comm})
and (\ref{mom-int}).


%

\paragraph{Adding twisted hypers}
To obtain the most general model, one should also add twisted hyper-multiplets.  One simply introduces
another target space $\tilde{X}$ for the twisted hypers and define
the corresponding moment map multiplet and so on. One also defines
\be
\mu^{mn} = -g_{ij} (V^m)^i (V^n)^j \,,
\;\;\;
\tilde{\mu}^{mn} = -\tilde{g}_{ij} (\tilde{V}^m)^i (\tilde{V}^n)^j \,.
\label{mumu2}
\ee
The minus sign is required for (\ref{mumu2}) to reduce to (\ref{mumu1})
in the linear case.

\paragraph{Lagrangian and supersymmetry}
Using the notations introduced so far,
we can write down the Lagrangian for the most general
Chern-Simons gauged non-linear sigma model:
\begin{eqnarray}
 \CL &=&
\CL_{\rm CS}(A)
\nn\\&&
 +\omega_{AB}
  \left(-\epsilon^{\a\b} D q^A_\a D q^B_\b
        +i\epsilon^{\da\db}\psi^A_\da \Ds\psi^B_\db \right)
 + \widetilde{\omega}_{AB}
  \left(-\epsilon^{\da\db} D \tilde q^A_\da D \tilde q^B_\db
       +i\epsilon^{\a\b}\tilde\psi^A_\a \Ds\tilde\psi^B_\b \right)
 \nn\\&&
 -i k_{mn} \left( \e^{\a\b} \e^{\dg\dd}\jmath^m_{\a\dg}\jmath^n_{\b\dd}
 + \e^{\da\db} \e^{\g\d}\tilde\jmath^m_{\da\g}\tilde\jmath^n_{\db\d}
 +4  \epsilon^{\a\g}\epsilon^{\db\dd}
   \jmath^m_{\a\db}\tilde\jmath^n_{\dd\g}
   -\epsilon^{\da\dg}\epsilon^{\db\dd}
 \tilde\mu^m_{\da\db} \rho^n_{\dg\dd}
 -\epsilon^{\a\g}\epsilon^{\b\d}
  \mu^m_{\a\b}\tilde{\rho}^n_{\g\d}  \right)
 \nn\\&&
  -\frac{1}{12} f_{mnp} \left( (\mu^m)^\a_{~\b}(\mu^n)^\b_{~\g}(\mu^p)^\g_{~\a}
   +(\tilde\mu^m)^{\dot{\a}}_{~\dot{\b}}
   (\tilde\mu^n)^{\dot{\b}}_{~\dot{\g}}
   (\tilde\mu^p)^{\dot{\g}}_{~\dot{\a}} \right)
 \nn\\&&
  -\half  \tilde\mu^{mn} (\mu_m)^\a_{~\b}(\mu_n)^\b_{~\a}
  -\half  \mu^{mn}  (\tilde\mu_m)^\da_{~\db}(\tilde\mu_n)^\db_{~\da}
  \nn \\
&& + \frac{1}{6} \O_{ABCD} (\psi^A_\da \psi^B_\db)(\psi^C_\dg \psi^D_\dd) \e^{\da\dg}\e^{\db\dd} + \frac{1}{6} \widetilde{\O}_{ABCD} (\tps^A_\a \tps^B_\b)(\tps^C_\g \tps^D_\d) \e^{\a\g}\e^{\b\d}  \,.
\label{Lful5}
\end{eqnarray}
The supersymmetry transformation rules read
\begin{eqnarray}
 && (\delta_\eta q)^{A\a}
    = i\eta^{\a\da}\psi_{\dot\alpha}^A\,,
~~~
    (\delta_\eta \tilde q)^{A\da}
    = i\eta^{\a\da}\tilde\psi_\alpha^A\,,
~~~
    \delta_\eta A^m_\mu =  i\eta^{\alpha\dot\alpha}\gamma_\mu
            (\jmath^m_{\alpha\dot\alpha}+\tilde\jmath^m_{\dot\alpha\alpha})\,,
 \nn\\
 && \delta_\eta \psi_{\dot\alpha}^A
    = - \left[(\sla D q)^{A\a}
           +\frac{1}{3} (V^m)^{A\b} (\m_m)^\a{}_\b \right]
            \eta_{\a\da}
           -  (V^m)^{A\b} (\tilde\mu_m)^{\dot\beta}_{~\dot\alpha}
            \eta_{\b\db}
            -  \G_i{}^A{}_B (\d_\eta q^i) \psi^B_\da \,,
 \nn\\
 && \delta_\eta \tps^A_\a
    = - \left[(\sla D \tq)^{A\da}
           +\frac{1}{3} (\tilde{V}^m)^{A\db} (\tilde{\m}_m)^\da{}_\db \right]
            \eta_{\a\da}
           -  (\tilde{V}^m)^{A\db} (\mu_m)^\b{}_\a
            \eta_{\b\db} -  \tilde{\G}_i{}^A{}_B (\d_\eta \tq^i) \tps^B_\a \,.
            \nn \\
            \label{Sful5}
\end{eqnarray}
Here,
we used some short-hand notations such as $Dq^{A\a} \equiv e^{A\a}_i Dq^i$ and
$(\d_\eta q)^{A\a} \equiv e^{A\a}_i (\d_\eta q)^i$.
The parameter $\eta$ satisfies the reality condition
\be
\label{eta-real2}
(\eta^{\a\da})^*
= + \e_{\a\b} \e_{\da\db} \eta^{\b\db} \,.
\ee

\paragraph{Mass deformation}
Unlike the linear model of the previous subsection,
the non-linear model does not admit a mass deformation.
For the linear model, the $SU(2)_L \times SU(2)_R$ $R$-symmetry
acts on a target space as a tri-holomorphic Killing vector.
The mass deformed Lagrangian (\ref{d-lag}) and supersymmetry algebra (\ref{dsusy}) can be associated to this Killing vector in a standard way \cite{agf}. In the non-linear model, however, the $R$-symmetry
is no longer an isometry of the target space, so the mass deformation
is not allowed.

%

\paragraph{Examples}

Gaiotto-Witten \cite{gw} gave a classification of linear model
in terms of Lie super-algebra; see (\ref{lsa}).
The same classification can be used
even after adding twisted hypers,
and the resulting theory is typically a linear quiver
with product gauge group and alternating series of hypers and twisted hypers \cite{n4}.

Such a classification for non-linear model is not known,
but a method for generating non-linear models from
linear models was given in \cite{gw}.
The idea is to use a linear quiver allowed by the Lie super-algebra conditions.
The linear quiver has two open ends.
The non-linear model is obtained by taking the usual hyper-K\"aler quotient
\cite{hklr,kron} with all the gauge groups except those at the end points.

Another class of non-linear models was given by Kapustin-Saulina \cite{ks}.
The gauge group $G$  is any compact simple Lie group and the target space is
the cotangent bundle of the flag manifold, $T^*(G/T^r)$,
where $T^r$ is the maximal torus of $G$.

We do not have any new example to offer here. Instead, as an illustration, we present the simplest example $T^*(SU(2)/U(1)) = T^*(\mathbb{CP}^1)$ from Kapustin-Saulina \cite{ks} in our notation.
This space is endowed with the famous Eguchi-Hanson metric \cite{eh},
\be
ds^2 = f^{-2} dr^2 + \frac{r^2}{4}(\s_1^2 +\s_2^2 + f^2 \s_3^2)
\;\;\;\; (f = \sqrt{1-r^{-4}}) \,,
\ee
where $\s_a$ are left-invariant one-forms on $S^3$ satisfying
$d\s_1 = \s_2 \wedge \s_3\,$, etc. Explicitly, in the Euler-angle coordinate,
\be
\s_1 &=& +\sin\psi\, d\th + \cos\psi \sin\th\,d\phi \,,
\nn \\
\s_2 &=& -\cos\psi\, d\th + \sin\psi \sin\th\,d\phi \,,
\nn \\
\s_3 &=&  d\psi-\cos\th \,d\phi \,,
\ee
The vielbeins are written in terms of the invariant one-forms as
\be
&&
e^0 = f^{-1} dr, \;\;\;
e^1 = \thalf r \s_1, \;\;\;
e^2 = \thalf r \s_2, \;\;\;
e^3 = \thalf r f \s_3, \;\;\;\;\;
e^{A\a} = (e^0 \identity + i e^a \tau^a)^{A\a} \,,
\ee
where $\tau^a$ are the Pauli matrices.
The three symplectic forms are given by
\be
\O_{\a\b} = \O_a (\e \tau^a)_{\a\b} \,,
\;\;\; \O_{a} &=& e^0 \wedge e^a + \half \e^{abc} e^b \wedge e^c \,.
\ee
The gauge group $G=SU(2)$ acts on the target space as isometries $V^m$
leaving $\s_a$ invariant,
\be
V^1 &=& +\sin\phi\, \p_\th + \cot\th \cos\phi\,\p_\phi +\csc\th \cos\phi\,\p_\psi\,,
\nn \\
V^2 &=& -\cos\phi\, \p_\th + \cot\th \sin\phi\,\p_\phi +\csc\th \sin\phi\,\p_\psi\,,
\nn \\
V^3 &=&  -\p_\phi \,,
\ee
The moment maps can be computed by solving (\ref{moment-map}): $i_{V^m} \O_a = d \mu^m_a$.
Checking the fundamental identity, we find
\be
k_{mn} \m^m{}_a \m^n{}_b =
\begin{pmatrix}
r^4-1 & 0 & 0 \\
0 & r^4-1& 0 \\
0 & 0 & r^4
\end{pmatrix}_{ab} \,.
\ee
Removing the trace part, we see that the strong version of fundamental identity
(\ref{fundi2}) does not hold. However, since the discrepancy
is a constant, the weaker version (\ref{fundi3}) holds
and the theory is well-defined.

All known examples so far, those of Gaiotto-Witten \cite{gw} and of Kapustin-Saulina \cite{ks} are non-compact. It is not clear (to us)
whether there exists any compact hyper-K\"ahler manifold
satisfying either the strong or the weak version of fundamental identity.


\section{Topological twisting}

\subsection{Survey of possible twistings}  \label{twisting}

Before performing the topological twisting of the
Chern-Simons sigma model described in the last section,
we pause to enumerate physically inequivalent twistings
for $\CN = 4,5,6,8$ theories.
A similar discussion is well-known in four dimensions,
where there is only one twisting for $\CN=2$ super-Yang-Mills
\cite{w88}
and three different twistings for $\CN=4$ super-Yang-Mills
\cite{yam,vw,marc}.
\footnote{
We will not consider the possibility of including conformal supercharges
in twisting \cite{park}.}

\paragraph{$\CN=4$ .}

The supercharges transform in the $({\bf 2,2})$ representation
under the $SO(4)\simeq SU(2)_L\times SU(2)_R$ $R$-symmetry group.
Recall also how the matter fields behave under the $R$-symmetry:
\be
q : ({\bf 2,1}), \;\;\;
\psi : ({\bf 1,2}), \;\;\;
\tilde{q} : ({\bf 1,2}), \;\;\;
\tilde{\psi} : ({\bf 2,1}).
\ee
Consider a theory with hyper-multiplets $(q,\psi)$ only.
Twisting with $SU(2)_R$ gives scalar supercharges in doublet of $SU(2)_L$
and matter fields transforming in $SU(2)_L \times SU(2)_{E'}$ ($SU(2)_{E'} = {\rm diag}\left[SU(2)_E \times SU(2)_R\right]$) as
\be
q : ({\bf 2,1}), \;\;\; \psi : ({\bf 1,1})\oplus ({\bf 1,3}).
\ee
We will call this ``A-twisting.''
Switching the roles of $SU(2)_L$ and $SU(2)_R$
leads to a different twisting, which we call ``B-twisting.''
Equivalently, we can continue to use $SU(2)_R$ for
twisting and consider a theory with twisted hyper-multiplets only.
In the latter convention, the resulting theory contains matter fields
transforming in $SU(2)_L \times SU(2)_{E'}$ as
\be
\tilde{q} : ({\bf 1,2}), \;\;\; \psi : ({\bf 2,2}).
\ee

We will continue to use $SU(2)_R$ for twisting even when both types of hyper-multiplets are present; the other twisting amounts to exchanging the roles of hyper and twisted hyper-multiplets.
We will call this broader class of theories ``AB-models,"
to distinguish them from the two extreme cases.
In general, $\CN=4$ theories are not symmetric under the ``mirror reflection" between hyper and twisted hyper-multiplets, while $\CN>4$ theories
are automatically symmetric.

\paragraph{$\CN=5$ .}

The supercharges transform as vector ${\bf 5}$ under the $SO(5)$ $R$-symmetry group, which have the following decomposition:
\begin{eqnarray}
  \text{case 1} &:&
  SO(5) \ \to \  SO(4) \simeq SU(2)_1 \times SU(2)_2 \,, \nn \\
  \text{case 2} &:&
  SO(5)  \ \to \ SO(3) \times SO(2)  \simeq SU(2)_3 \times U(1) \,.\nn
\end{eqnarray}
The case 1 is the same as the AB-twisting of $\CN=4$ theories.
In the case 2, the supercharges transform as ${\bf 3}_0 \oplus {\bf 1}_+ \oplus {\bf 1}_-$
and do not yield any scalar supercharge upon twisting.

\paragraph{$\CN=6$ .}

The supercharges transform as vector ${\bf 6}$ under the $SO(6)$ $R$-symmetry group, which have the following decomposition:
\begin{eqnarray}
  \text{case 1} &:&
  SO(6) \ \to \  SO(4)\times SO(2) \simeq SU(2)_1 \times SU(2)_2 \times U(1) \,, \nn \\
  \text{case 2} &:&
  SO(6)  \ \to \ SO(3) \times SO(3)  \simeq SU(2)_3  \times SU(2)_4 \,. \nn
\end{eqnarray}
The case 1 can be regarded as a refinement of the AB-twisting of $\CN=4$
theories in the sense that there is a left-over $U(1)$ $R$-symmetry group.
The scalar supercharge is neutral under this $U(1)$, but
the theory contains charged matter fields.
The case 2 can be regarded as a refinement of the case 2 of $\CN=5$ theories
considered above. Again, we find that there is no possible twisting to have scalar supercharges.

\paragraph{$\CN=8$ .}

The $\CN=8$ BLG theories have the $R$-symmetry group $SO(8)$.
If we keep the supercharges in the vector $\mathbf{8}_v$ representation,
we will only obtain refinements of the AB-twisting of $\CN=4$ theories.
New possibilities may arise if we use the triality of $SO(8)$ to
let the supercharges transform in the spinor $\mathbf{8}_s$ representation.

Consider the following decompositions:
\begin{eqnarray}
\text{case 1} &:&
  SO(8)   \to  SO(2) \times SO(6)   \to
  SO(2) \times SO(3)\times SO(3)  \simeq
  U(1) \times SU(2)_3 \times SU(2)_4
  \nonumber \\
\text{case 2} &:&
  SO(8)   \to  SO(2) \times SO(6)   \to
  SO(2)^2 \times SO(4)  \simeq  U(1)^2 \times SU(2)_1 \times SU(2)_2
  \nonumber \\
\text{case 3} &:&
  SO(8)   \to  SO(3) \times SO(5)  \simeq SU(2)_A  \times USp(4)
  \nonumber \\
\text{case 4} &:&
  SO(8)  \to   SO(4)\times SO(4) \simeq SU(2)_a \times SU(2)_b \times SU(2)_c
  \times SU(2)_d \nonumber
  \end{eqnarray}
Notice that the cases 1 and 2 can be enhanced to the cases 3 and 4, respectively.
So, it is sufficient to examine the latter two cases.

Under the subgroups of case 3, the supercharges transform as $(\bf{2,4})$.
It appears that we have one possible twisting with four scalar-supercharges
in the ${\bf 4}$ representation of $SO(5)\simeq USp(4)$. However, there is a slight subtlety here.
Since the triality of $SO(8)$ is broken in the decomposition,
we have two choices for the representation of the matter fields,
which leads to two inequivalent twistings. Denoting the scalar
and fermion fields by $\Phi$ and $\Psi$ and specifying how the representations
of the $SU(2)_E\times SU(2)_A \times USp(4)$ before the twisting
reduces to those of the $SU(2)_{E'} \times USp(4)$ after twisting, we find
\begin{itemize}

\item
C-twisting :
\be
\Phi &:& ({\bf 1;3,1})\oplus ({\bf 1;1,5}) \ \to \ ({\bf 3;1})\oplus ({\bf 1,5})
\,, \nn \\
\Psi &:& ({\bf 2;2,4})\oplus \to \ ({\bf 3;4})\oplus ({\bf 1;4}) \,. \nn
\ee

\item
D-twisting :
\be
\Phi &:& ({\bf 1;2,4}) \ \to \ ({\bf 2;4})
\,, \nn \\
\Psi &:& ({\bf 2;3,1})\oplus ({\bf 2;1,5}) \ \to \ ({\bf 4;1})\oplus ({\bf 2;1})
\oplus ({\bf 2;5}) \,. \nn
\ee

\end{itemize}

In the case 4, the triality of $SO(8)$ survives as
permutations of the four $SU(2)$ factors;
${\bf 8}_v$, ${\bf 8}_s$ and ${\bf 8}_c$ are all related by permutations. 
We can see it from the following assignment for the supercharges and matter fields,
\begin{eqnarray}
  Q &:& ( {\bf 2,1,2,1}) \oplus ( {\bf 1,2,1,2} )\ ,
  \nn \\
  \Phi &:& ( {\bf 2,2,1,1}) \oplus ( {\bf 1,1,2,2} )\ ,
  \nn \\
  \Psi &:& ( {\bf 2,1,1,2}) \oplus ( {\bf 1,2,2,1} )\ .
\end{eqnarray}
There is a novelty here. Unlike all other cases considered so far, we can now
use more than one $SU(2)$ factors for twisting.
Up to permutations, we have four candidates for different twistings.

\begin{itemize}

\item
AB, further refined :
${\rm diag}[SU(2)_E \times SU(2)_a] \times SU(2)_b\times SU(2)_c \times SU(2)_d$,
\be
Q &:& ( {\bf 1;1,2,1}) \ ,
  \nn \\
  \Phi &:& ({\bf 2;2,1,1}) \oplus ( {\bf 1;1,2,2} )\ ,
  \nn \\
  \Psi &:& ( {\bf 3;1,1,2}) \oplus( {\bf 1;1,1,2}) \oplus ( {\bf 2;2,2,1} )\ .
\ee

\item
C'-twisting : $SU(2)_{E'} = {\rm diag}[SU(2)_E \times SU(2)_a \times SU(2)_b]\times SU(2)_c \times SU(2)_d$,
\be
Q &:& ( {\bf 1;2,1})  \oplus ( {\bf 1;1,2}) \ ,
  \nn \\
  \Phi &:& ({\bf 3;1,1}) \oplus ( {\bf 1;1,1} )\oplus ( {\bf 1;2,2} )\ ,
  \nn \\
  \Psi &:& ( {\bf 3;1,2}) \oplus( {\bf 1;1,2}) \oplus ( {\bf 3;2,1} )\oplus ( {\bf 1;2,1} )\ .
\ee

\item
D'-twisting : $SU(2)_{E'} = {\rm diag}[SU(2)_E \times SU(2)_a \times SU(2)_d]\times SU(2)_b \times SU(2)_c$,
\be
Q &:& ( {\bf 1;1,2})  \oplus ( {\bf 1;2,1}) \ ,
  \nn \\
  \Phi &:& ({\bf 2;2,1}) \oplus ( {\bf 2;1,2} )\ ,
  \nn \\
  \Psi &:& ( {\bf 4;1,1}) \oplus ( {\bf 2;1,1}) \oplus ( {\bf 2;1,1}) \oplus ( {\bf 2;2,2} ) \ .
\ee

\item
E-twisting : $SU(2)_{E'} = {\rm diag}[SU(2)_E \times SU(2)_a \times SU(2)_b \times SU(2)_c] \times SU(2)_d$,
\be
Q &:& ( {\bf 1;2}) \ ,
  \nn \\
  \Phi &:& ({\bf 3;1}) \oplus ( {\bf 1;1} )\oplus ( {\bf 2;2} )\ ,
  \nn \\
  \Psi &:& ( {\bf 3;2}) \oplus ( {\bf 1;2}) \oplus ( {\bf 4;1}) \oplus ( {\bf 2;1} )\oplus ( {\bf 2;1} ) \ .
\ee

\end{itemize}

\noindent
Note that the C' and D' twistings are not really new
as they can be obtained from the C and D twistings, respectively,
by breaking $USp(4)$ down to $SU(2)\times SU(2)$.

The C-twisting of the BLG theory has been studied in \cite{Lee:2008cr}.
In the rest of this section,
we will construct the A/B/AB-twisted
Chern-Simons sigma models, leaving
the other twistings for a future work.



\subsection{A-model}

We begin with the Euclidean version of the Lagrangian
of the linear-model,
\begin{eqnarray}
 \CL &=&
 -i \CL_{\rm CS}(A)
 + \omega_{AB}
  \left(\epsilon^{\a\b} D q^A_\a D q^B_\b - i\epsilon^{\da\db}\psi^A_\da \Ds\psi^B_\db \right)
 \nn\\&&
 + i k_{mn} \e^{\a\b} \e^{\dg\dd} \jmath^m_{\a\dg} \jmath^n_{\b\dd}  + \frac{1}{12} f_{mnp} (\mu^m)^\a_{~\b}(\mu^n)^\b_{~\g}(\mu^p)^\g_{~\a} \,.
\label{Lful-Euc1}
\end{eqnarray}
This Euclidean Lagrangian differs from the Lorentzian one (\ref{Lful1})
by the factor of $(-i)$ for the Chern-Simons term
and an overall sign for the matter terms.
Besides, the fermions no longer satisfy the reality condition,
but the Lagrangian depends holomorphically on the fermions.
The supersymmetry transformation rules remain formally the same
as in the Lorentzian theory,
\be
\label{susytr-Euc1}
 &&\delta q_\a^A = i\eta_\a{}^{\da} \psi_\da^A \,,~~~~
 \delta A^m_\m =  i \eta^{\a\da} \gamma_\m \jmath^m_{\a\da}\,,
  \nn \\
 &&\delta\psi_\da^A=
 \left[\Ds q_\a^A + \frac{1}{3}k_{mn}(t^m)^A_{~B} q^B_\b (\mu^n)^\b_{~\a}\right]
 \eta^\a_{~\;\da}\,.
\ee

\paragraph{Twisting}
The twisting is done by taking the diagonal subgroup of the Lorentz group $SU(2)_E$ and the $SU(2)_R$ part of the $R$-symmetry group as the new Lorentz group: $SU(2)'_E = {\rm diag}[ SU(2)_E \times SU(2)_R ]$. For instance, we
make the replacements,
\begin{eqnarray}
\label{twist1}
(\psi_\s)^A_\da = \frac{1}{\sqrt{2}} \left( i \lambda^A \epsilon_{\s \da} + \chi_{ \m}^A  ( \gamma^{ \m } \e)_{\s \da}  \right) \,,
  \;\;\;
(\eta^\s)_\a{}^\da  = \frac{1}{\sqrt{2}} \eta_\a \e^{\s\da}
 \,,
\end{eqnarray}
where $\s$ denote the Lorentz indices, while $(\a, \da)$ denotes two $SU(2)$ $R$-symmetries indices.
We suppressed the $SU(2)'_E$-triplet components of the parameter $\eta$.
%
%
%
%
%
%

Plugging these into the transformation rules (\ref{susytr1}),
we find the following twisted transformation rules:
\begin{eqnarray}
\label{twi-susy1}
  \d_\eta q^A_\a = \eta_\a \lambda^A \,, \hspace{0.5cm}
  \d_\eta A_{ \m}^m = -i \eta^\a (j^m_{ \m })_{\a } \,, \hspace{0.5cm}
  \d_\eta \lambda^A = i H^A_\a \eta^\a, \hspace{0.5cm}
  \d_\eta \chi^A_{ \m }  = - D_{ \m }  q^A_\a \eta^\a\  ,
\end{eqnarray}
where we defined
\be
\, \;\;\
( j^m_{ \sigma} )_{ \a \da} \equiv \frac{1}{ \sqrt{2} } ( i \epsilon_{ \sigma \da} j^m_{ \a} + ( \gamma^{ \m }  \epsilon)_{ \s \da} ( j^m_{ \m }  )_{ \a } )\, , \quad
H^A_\a \equiv  \frac{1}{3} (t_m)^A_{\ B} q^B_\b (\mu^m)^\b_{\ \a}\,.
\label{def:F}
\ee
{}From the definition, one can show that
\be (j^m_{ \m } )_{ \a } = t^m_{ AB} q^A_{ \a} \chi_{ \m }^B \, ,
\quad (j^m)_{ \a} = t^m_{ AB} q^A_{ \a} \lambda^B \, . \ee
Introducing the twisted supercharges by $\d_\eta X = [ \eta^\a Q_\a, X]$, we can rewrite (\ref{twi-susy1}) as
 \begin{eqnarray}
 \label{twi-susy2}
  {}[ Q_\a , q_\b^A ] = - \epsilon_{\a\b}\lambda^A, \hspace{0.15cm}
  {}[ Q_\a, A_{ \m }^m ] =  -i(j^m_{ \m })_{ \a }  , \hspace{0.15cm}
  \{ Q_\a , \lambda^A \} = iH^A_\a , \hspace{0.15cm}
  \{ Q_\a , \chi_{ \m }^A \} = -  D_{ \m }  q^A_\a .
\end{eqnarray}

\paragraph{Nilpotency}

The twisted supercharges $Q_\a$ are nilpotent up to a gauge transformation.
Concretely, the following relations hold,
\be
\label{nil1}
\left[  \{ Q_\a , Q_\b \}, q^A_\g \right] = - \left( \L_{\a\b} \right)^A_{\ B} q^B_\g \,,
&&
\left[ \{Q_\a , Q_\b\}, A_{ \m }^m \right] = D_{ \m }  \L_{\a\b}^m \,,
\nn \\
\left[ \{ Q_\a , Q_\b \}, \lambda^A \right]
= - (\L_{\a \b})^A_{\ B} \lambda^B \,,
&&
\left[ \{ Q_\a, Q_\b\} ,\chi^A_{ \m }  \right] = - ( \Lambda_{ \alpha \beta} )^A_{ \ B } \chi_{ \m }^A  \,,
\ee
with the gauge parameter
\be
\label{Lambda}
 \Lambda_{ \a \b}^m = i \mu^m_{ \a \b} \ ,
\qquad
(\L_{\a \b})^A_{\ B}  \equiv   (t_m)^A_{\ B} \Lambda^m_{\a\b} \,.
\ee
The fundamental identity $k_{mn} t^m_{(AB}t^n_{C)D}=0$
is needed to verify (\ref{nil1}).
See appendix \ref{app:detail} for details.


\paragraph{Lagrangian and Invariance}

The twisted Lagrangian can be divided into two pieces, 
\be
\CL &=& \CL_1 + \CL_2 \,,
\\
\CL_1 &=& - i\CL_{\rm CS}(A) + \ve^{ \m \n \r } \w_{AB} \chi_{ \m }^A D_{ \n }  \chi_{ \r }^B \,,
\label{A_CL1} \\
\CL_2 &=& \omega_{AB} \epsilon^{\a\b} D q^A_\a D q^B_\b - 2i \omega_{AB} \lambda^A D^{ \m }  \chi^B_{ \m }
+ i k_{mn} \e^{\a\b}  ( ( \jmath^m_{\m} )_{\a } ( \jmath^{ \m n} )_{ \b }  + \jmath^m_{\a} \jmath^n_{\b} )
\nn \\
&& + \frac{1}{12} f_{mnp} (\mu^m)^\a_{~\b}(\mu^n)^\b_{~\g}(\mu^p)^\g_{~\a}\ .
\label{A_CL2}
\ee
The two parts $\CL_1$ and $\CL_2$ are $Q$-invariant separately.
The $Q$-invariance of $\CL_1$ can be checked explicitly,
\begin{eqnarray}
  {}[Q_\a , \CL_1 ] =  -  \e^{ \m \n \r } k_{mn}
  (q^A_\a t^m_{AB} \chi^B_{ \m } ) F^n_{ \n \r } +  \omega_{AB} \e^{ \m \n \r } q^A_\a
  [D_{ \m } ,D_{ \n } ]   \chi^B_{ \r }  +\cdots = 0 \,,
\end{eqnarray}
where $ \cdots $ denotes a part which vanishes due to the fundamental identity.
$\CL_2$ is also $Q$-invariant because it is $Q$-exact in the sense that
\be
\left\{ Q_\a , 2 \w_{AB}(\chi^A_{ \mu}  D^{ \mu }  q^B_\b -  i H^A_\b \lambda^B ) \right\}
&=&   \e_{\a\b} \CL_2  \,.
\ee

\paragraph{Mass deformation}
For the mass deformed linear model,
the A-twisting leads to the following super-algebra
for the scalar super-charges $Q_\a$,
\be
\label{dsusy2}
  \left\{ Q_{\a} , Q_{\b} \right\} \sim R_{\a\b}\ ,
\ee
where $R_{\a\b}$ denote the generators of $SU(2)_L$; see (\ref{dsusy}).
Without nilpotent scalar super-charges, we cannot define a topological field theory. Note also that (\ref{dsusy2}) somewhat resembles,
but clearly differs from, the defining relation of equivariant cohomology
which states
\be
Q^2 \sim k^{\a\b} R_{\a\b} \,,
\ee
for some parameters $k^{\a\b}$.
As a side remark, we note that $\CN=2$ super Yang-Mills theory in four dimensions
admit a mass deformation of a different kind if
the world-volume is a K\"ahler manifold \cite{w94}.

\paragraph{Non-linear A-model}

We now consider applying the A-twist to the non-linear model discussed in section \ref{non-linear}. 
The twisted transformation rules are as follows,
\be && ( \delta_{ \eta} q )^{ A \a } = \eta^{ \a }  \lambda^A, \quad \delta_ \eta A_{ \mu}^m  = - i \eta^{ \a } ( j^m_{ \mu} )_{ \a } , \nn \\
&& \delta_{ \eta} \lambda^A =  H^A_{ \a } \eta^{ \a} - \Gamma_{ i \ \  B}^{ \ A } ( \delta_{ \eta } q^i ) \lambda^B, \quad \delta_{ \eta} \chi_{ \mu}^A = ( D_{ \mu} q )^{ A \a } \eta_{ \alpha } - \Gamma_{ i \  \ B }^{ \ A} ( \delta_{ \eta} q^i ) \chi_{ \mu}^B  ,
\label{susy-nlA}
\ee
where the definitions of $H^A_{ \a } , ( j^m_{ \mu} )_{ \a } $ are appropriately covariantized in terms of the Killing vectors of the target space,
\be H^A_{ \a } \to \frac{1}{3} k_{ mn} ( V^m)^{ A \b } ( \mu^n )_{ \a \b } , \quad (j^m_{ \mu} )_{ \a }  \to  - V^m_{ A \a } \chi_{ \mu}^A  .
\ee
The main difference from the linear model is the four fermion term inherited from the physical theory,
\be \CL_1 & = & - i \CL_{\rm CS}(A) + \varepsilon^{ \mu \nu \rho } \left(  \omega_{ AB} \chi_{ \mu}^A D_{ \nu} \chi_{ \rho }^B + \frac{1}{3} \Omega_{ ABCD} \chi_{ \mu}^A \chi_{ \nu}^B \chi_{ \rho}^C \lambda^D \right) .  \nn
\ee
The $Q$-exact part, $\CL_2$, becomes a covariantized version of (\ref{A_CL2}).

Let us remark on differences from the Rozansky-Witten theory in \cite{rw}. Firstly, the derivatives become gauge covariant, thus the variation of the fermion kinetic term can cancel the variation of the Chern-Simons term.
Secondly, $ \delta_{ \eta }  \lambda^A = 0$  in the RW theory, while  $ \delta_{ \eta} \lambda^A \neq 0 $ in \eqref{susy-nlA}
due to the non-trivial bosonic potential. It suggests that the variation of $ \lambda$ of the curvature term shall be canceled by the variation of the gauge boson in the fermionic kinetic term. The cancellation has been shown in detail in \cite{ks} for a holomorphic supercharge.

\subsection{B-model}
Next, we consider a theory only with twisted hyper multiplets. The physical Lagrangian in the Euclidean signature is
\begin{eqnarray}
 \CL &=& - i \CL_{\rm CS}(A)
 + \omega_{AB}
  \left(\epsilon^{\da\db} D \tilde q^A_\da D \tilde q^B_\db
       - i\epsilon^{\a\b}\tilde\psi^A_\a \Ds\tilde\psi^B_\b \right)
 \nn\\&&
 + i k_{mn}
 \e^{\da\db} \e^{\g\d}\tilde\jmath^m_{\da\g}\tilde\jmath^n_{\db\d}
  +\frac{1}{12} f_{mnp}
   (\tilde\mu^m)^{\dot{\a}}_{~\dot{\b}}
   (\tilde\mu^n)^{\dot{\b}}_{~\dot{\g}}
   (\tilde\mu^p)^{\dot{\g}}_{~\dot{\a}}
 \label{Lful-B}
\end{eqnarray}

\paragraph{Twisting}
 $ \tilde{q}^A_{ \da } $ becomes a bosonic spinor in $ ( { \bf 2} ; { \bf 1} ) $ representation of $su(2)_{ E}^{ \prime} \oplus su(2)_L $.  For later convenience, we rescale $ \tilde{ \psi}^A_{ \a } $, which is in $({ \bf 2} ; { \bf2})$,  as
 \be ( \tilde{ \zeta}_{ \sigma} )^A_{ \a } \equiv \sqrt{2} ( \tilde{ \psi}_{ \sigma} )^A_{ \a } \ .  \ee
The supersymmetry variations become
\be\label{Btransf}
[ Q_{  \a } , \tilde{q}^A ] = - \frac{i}{2} \tilde{ \zeta }^A_{ \a } , \quad [ Q_{  \a } , A_{ \m }^m ] = - i ( \tilde{j}^m_{ \m } )_{  \a } ,
 \quad  \{ Q_{ \a } , \tilde{ \zeta }^A_{ \b } \} = - \epsilon_{  \a  \b } ( \Ds \tilde{q}^A + \tilde{H}^A )  \ . \ee
Note that spinor indices of $su(2)_E^{ \prime}$ are implicit.
The quantities $ \tilde{j}^m_\m $ and $\tilde{H}^A$ are defined
in the same way as \eqref{def:F}.

\paragraph{Nilpotency}
The twisted supercharges $Q_{  \a }$ are nilpotent up to the equations of motion of the fermion.
The results read
\be
 [ \{ Q_{  \a  } , Q_{ \b } \},  \tilde{q}^A ] = 0 \ , \quad [ \{ Q_{ \a  } , Q_{  \b } \},  A_{  \m }^m ] = 0  \ , \quad   [ \{ Q_{  \a } , Q_{  \b } \},  \tilde{ \zeta }^A_{  \g } ] =  \epsilon_{  \b  \g }  \xi^A_{  \a } + \epsilon_{  \a \g } \xi^A_{  \b } \ ,
\label{on-shell}
\ee
where
\be \xi^A_{ \a } \equiv  \frac{i}{2}   \left( \Ds \tilde{ \zeta }^A_{ \a } + \gamma^{ \m } ( \tilde{ j}^m_{ \m } )_{  \a } + i (\tilde{t}^m)^A_{ \ B } \tilde{q}^B ( \tilde{j}_m )_{  \a } \right) \  .
\ee
Note that the variation of the Lagrangian with respect to the fermion is,
up to a surface term,
\be
\delta \CL = - 4 \omega_{ AB} \epsilon^{ \a \b} ( \delta \tilde{ \zeta }^A_{ \a} ) \xi^B_{ \b} \,.
\ee

An aspect of \eqref{on-shell} different from the result of A-model  \eqref{nil1} is the absence of the gauge transformation on the right hand side. One can expect it from the symmetry, sine if  a field-dependent gauge transformation parameter exists, it should be in $  ( { \bf 1} ;{ \bf 3} )$ representation of  $su(2)_E^{ \prime} \oplus su(2)_L $ with conformal dimension 1.
A short computation shows that such a composite field cannot be constructed from
the matter fields of the B-model.

\paragraph{Off-shell Supersymmetry Algebra}
One can construct an off-shell formalism
by introducing an auxiliary field, a bosonic spinor $ \tilde{h}^A$. The off-shell variations are defined as follows
\be
 \{ Q_{  \a }, \tilde{ \zeta }^A_{ \b } \} = - \epsilon_{  \a   \b } ( \Ds \tilde{q}^A + \tilde{h}^A ) \ ,  \quad [ Q_{  \a } ,  \tilde{h}^A ] = \frac{i}{2} \Ds \tilde{ \zeta }^A_{  \a } + i   \gamma^{ \m }  ( \tilde{ j}^m_{ \m } )_{  \a  } ( \tilde{t}_m )^A_{ \ B }  \tilde{q}^B \ .
\label{off-shell}
 \ee
The variations of $ \tilde{q}^A$ and $ A_i^m$ are unchanged from  \eqref{on-shell}. The variation of the auxiliary field in \eqref{off-shell} has been chosen to ensure the nilpotency of $Q_\a$ on $ \tilde{ \zeta }^A_{  \a } $, {\it i.e.}, $ [ \{ Q_{  \a } , Q_{  \b } \} , \tilde{ \zeta }^A_{  \g } ] = 0 $.  One can indeed check that
nilpotency holds off-shell for all fields $( \tilde{q}^A, A_\m^m, \tilde{ \zeta }^A_{ \dot{ \alpha } } ,  \tilde{h}^A ) $:
\be [ \{ Q_{  \a } , Q_{  \b }  \}, \, \cdot\, ] = 0 \ .  \ee

\paragraph{Off-shell Lagrangian}

The bosonic potential of the on-shell Lagrangian \eqref{Lful-B} can be rewritten in terms of $ \tilde{H}^A$,
\be \CL_V \equiv \frac{1}{12} f_{mnp} ( \tilde{ \mu}^m)^\a_{~\b}( \tilde{ \mu}^n)^\b_{~\g}( \tilde{ \mu}^p)^\g_{~\a} \,
= - \omega_{ AB}   \tilde{H}^A  \tilde{H}^B  \  .
\ee
Given the off-shell supersymmetry algebra, the following replacement of $ \CL_V$ leads to the off-shell supersymmetric Lagrangian ,
\be   \CL_V  \to  \CL_V = \omega_{ AB} (-  \tilde{h}^A  \tilde{h}^B  + 2  \tilde{h}^A \tilde{H}^B - 2 \tilde{H}^A \tilde{H}^B ) \  .
\ee
%
Again, the twisted Lagrangian can be split into two pieces, $\CL=\CL_1+\CL_2$,
\begin{eqnarray}
 \CL_1 &=& - i \CL_{\rm CS}(A^+)  \ , \nn \\
 \CL_2 &=& -  \omega_{ AB} D \tilde{q}^A  D^{+} \tilde{q}^B  - \frac{i}{2} \omega_{AB} \epsilon^{  \a  \b } \tilde{ \zeta }^A_{ \a } \Ds^{+} \tilde{ \zeta }^B_{ \b }
+2 i k_{mn} \e^{\a\b}  ( ( \tilde{ \jmath}^m_{\m } )_{ \a } (  \tilde{ \jmath }^{n}_{ \m } )_{ \b } - \tilde{ \jmath }^m_{\a} \tilde{ \jmath }^n_{\b} )   \nn \\
&& \hspace*{-0.7cm}
+  i \omega_{ AB}  \varepsilon^{ \m \n \r } D_{ \m }   \tilde{q}^A \gamma_{ \n }  D^{+}_{ \r }   \tilde{q}^B  - \omega_{ AB} \tilde{H}^A \Ds^{+} \tilde{q}^B + \omega_{ AB} ( - \tilde{H}^A \tilde{H}^B + 2 \tilde{H}^A  \tilde{h}^B - \tilde{h}^A  \tilde{h}^B ) \ ,
\label{lag_B}
\end{eqnarray}
where the covariant derivative $D^+_\mu$ now involves $A^+_\mu$ rather than $A_\mu$.

The $Q$-closed Lagrangian $ \CL_1 $ is simply the Chern-Simons action, except
that the gauge boson is shifted by a bi-linear product of the boson fields,
\be (A^+)^m_\mu \equiv A^m_\mu + S^m_\mu \ ,
\qquad
{S}_{  \m }^m \equiv \half \tilde{ t}^m_{ AB} \tilde{q}^A \gamma_{ \m }  \tilde{q}^B \ .
\label{CA}
\ee
A similar redefinition of the gauge field has been noticed \cite{Lee:2008cr}
for the C-twisting of BLG theory discussed in section \ref{twisting}
and even earlier in topological Yang-Mills theories in \cite{marc,bth,park}.
The meaning of this shift will be discussed in section \ref{last}.

As in the A-model, $\CL_1$ and $ \CL_2$ are separately $Q$-invariant. $Q$-invariance of $ \CL_1 $ is trivial due to the $Q$-invariance of $ { A^+_{  \m } }^m$,
\be [ Q_{ \a } , { A^+_{ \m } }^m ] = 0 \ . \ee
$ \CL_2$ is $Q$-exact,
\be \{ Q_{ \a }, \omega_{ AB} \left( - \tilde{ \zeta }^A_{  \b } (  \Ds^{ +}  \tilde{q}^B  -  \tilde{H}^B - \tilde{h}^B ) \right) \} = \epsilon_{  \a  \b } \CL_2 \ .
\ee

To show that the Lagrangian in \eqref{lag_B} indeed results from the topological twisting of the physical Lagrangian  \eqref{Lful-B}, one needs to use rather nontrivial identity
\be  k_{ mn} {S}^m_{ \m } ( \tilde{t}^n)_{ AB} \tilde{q}^A D^{ \m }  \tilde{q}^B = \omega_{ AB}  \tilde{H}^A \Ds \tilde{ q}^B \ .  \label{pf}  \ee
See appendix \ref{app:detail} for a proof.

\paragraph{Wilson-loop observable}
The BRST invariance of $A_\mu^{+m}$ implies that the Wilson-loop
operators $W(C)$ are good observables of the B-model,
\begin{eqnarray}
  W(C) = \text{tr}\left( \CP\exp{\oint_C A^+} \right)\ .
\end{eqnarray}
We will see in section \ref{last} that, not coincidentally, the B-twisted $\CN=4$ super Yang-Mills 
theory also carries a shifted gauge field and Wilson loop observables.

\paragraph{Non-linear B-model}
In the B-model, the bosonic matter fields become
world-volume spinors. As such, it seems difficult, if not impossible,
to generalize the model to non-flat hyper-K\"ahler target space.
A similar problem arise from the topological Seiberg-Witten theory
\cite{w94b} in four dimensions, in which the monopole fields are
bosonic spinors. In this case, a generalization to non-flat hyper-K\"ahler target space was proven possible \cite{af} by coupling
the hyper-K\"ahler structure of the four-dimensional world-volume
to that of the target space. Our B-model does not seem to allow for such a
construction.

\subsection{AB-model}


Finally, we move on to theories containing both hyper and twisted hyper multiplets. In addition to the linear combination of A and B-model, non-trivial mixing terms arise.
%
The supersymmetry transformation laws are unchanged from \eqref{Sful5}.

\paragraph{Twisting}

We begin with implementing the features of the B-model to the AB-model,
namely, the shift in the gauge field (\ref{CA}) and the introduction of
auxiliary fields (\ref{off-shell}).
Using the notations introduced in the previous subsections,
we can summarize the twisted supersymmetry variations as follows,
\be
&& [ Q_{ \a } , q^A_{ \b } ] = - \epsilon_{ \a \b} \lambda^A, \quad \{ Q_{ \a } , \lambda^A \} = i H^A_{ \a} , \quad \{ Q_{ \a} , \chi_{ \mu}^A \} = - {D^+_{ \m } } q^A_{ \a} ,  \nn \\
&& [ Q_{ \a} , \tilde{q}^A ] = - \frac{i}{2} \tilde{ \zeta}^A_{ \a} , \quad \{ Q_{ \alpha } , \tilde{ \zeta}^A_{ \b} \} = - \epsilon_{ \a \b} ( \Ds \tilde{q}^A + \tilde{h}^A ) + k_{ mn}  ( \tilde{t}^m)^A_{ \ B } ( \mu^n)_{ \a \b} \tilde{q}^B ,  \nn \\
&& [ Q_{ \a} , \tilde{h}^A ] = \frac{i}{2} \Ds \tilde{ \zeta}^A_{ \a} + i  \gamma^{ \m }  \left( (j^m_{ \m } )_{ \a} + ( \tilde{j}^m_{ \m } )_{ \a } \right) ( \tilde{t}_m )^A_{ \ B } \tilde{q}^B -  ( \tilde{t}^m)^A_{ \ B } (j_m)_{ \a} \tilde{q}^B - \frac{i}{2} ( \tilde{t}^m)^A_{ \ B } ( \mu_m )_{ \a \b } \zeta^{ B \b } \, ,  \nn \\
&& [ Q_{ \a} , A^m_{ \m }  ] = - i \left( (j^m_{ \m } )_{ \a} + ( \tilde{ j}^m_{ \m } )_{ \a} \right) \, ,
\ee
%
Again, the $Q$-variation of the auxiliary field $ \tilde{h}^A $ is chosen to guarantee the nilpotency on $ \tilde{ \zeta}^A_{ \a}$, in this case up to a gauge transformation to be discussed next.

\paragraph{Nilpotency}
Now we deal with the problem inherited from the A-model; the twisted super-charges are nilpotent up to a gauge transformation. For  $(q^A_{ \g } , \lambda^A, \chi_{ \m }^A )$ in the hyper multiplet, the results in
\eqref{nil1} still hold, with the gauge parameter $ \Lambda_{ \a \b}^m $ in \eqref{Lambda}. Consistently,  $Q^2$ acts on the twisted hyper-multiplet $( \tilde{q}^A , \tilde{ \zeta}^A_{ \g} , \tilde{h}^A)$ as well as the modified gauge field ${ A^+_{ \m } }^m$ as a gauge transformation by the same gauge parameter,
\be
&& [ \{ Q_{ \a} , Q_{ \b} \} , \tilde{q}^A ] = - ( \tilde{ \Lambda}_{ \a \b} )^A_{ \ B } \tilde{q}^A \ , \quad [ \{ Q_{ \a } , Q_{ \b } \} , \tilde{ \zeta}^A_{ \g } ] = - ( \tilde{ \Lambda}_{ \a \b} )^A_{ \ B } \tilde{ \zeta}^B_{ \g} \ ,  \nn \\
&& [ \{ Q_{ \a } , Q_{ \b } \} , \tilde{ \CF}^A ] = - ( \tilde{ \Lambda}_{ \a \b }  )^A_{ \ B } \tilde{ \CF}^B, \quad   [ \{ Q_{ \a} , Q_{ \b} \} , { A^+}^m ] =  D^+ \Lambda^m_{ \a \b} \ .
\label{nil:AB}
\ee
where
\be ( \tilde{ \Lambda}_{ \a \b} )^A_{ \ B } \equiv \Lambda^m_{ \a \b} ( \tilde{t}_m )^A_{ \ B } \, .
\ee

\paragraph{Lagrangian and Invariance}
The Lagrangian still admits the usual splitting,
$ \CL  = \CL_1 + \CL_2  $.
The $\CL_1$ term is almost the same as in the A-model,
%
\be
\CL_1 & = & - i \varepsilon^{ \m \n \r } \left( k_{ mn} { A^+_{ \m } }^m \p_{ \n } { A^+_{ \r } }^n + \frac{1}{3} f_{ mnp} { A^+_{ \m }  }^m { A^+_{ \n } }^n { A^+_{ \r } }^p  + i \chi^A_{ \m } D^+_{ \n }  \chi^B_{ \r }   \right) \, ,  \label{AB_CL1}
\ee
except for the shift from $A^m$ to ${A^m}^+$ we noticed in the B-model.
%
The cubic term of the shift $S^m \equiv {A^m}^+ - A^m$ in the Chern-Simons term comes from the bosonic potential $ \tilde{ \mu}^3$ in the physical Lagrangian, while the change in the covariant derivative $ D \chi \to D^{+} \chi $ has its origin in the Yukawa term, $ \tilde{ \mu}^m \rho_m $.
It is straightforward to show that $ \CL_1 $  is $Q$-invariant,
\be [ Q_{ \a } , \CL_1 ]  = - \varepsilon^{ \m \n \r } k_{ mn} {F^+_{ \m \n } }^m (j^n_{ \r } )_{ \a } + \varepsilon^{ \m \n \r } \chi_{\m}^A [ D^+_{ \n } , D^+_{ \r }  ] q^B_{ \a }   \ = 0  \, ,   \ee
where $ {F^+_{ \m \n } }^m  \equiv [ D^+_{ \m } , D^+_{ \n }  ]^m $ is the field strength of $ { A^+_\m }^m $.
$ \CL_2$ is $Q$-exact in the sense that
\be \{ Q_{ \a} , f^{+}_{ \b} +2 i ( \mu_m)_{ \b}^{ \ \g} ( \tilde{j}^m)_{ \g } - 4 \left( (j^m_\m + \tilde{j}^m_{ \m } \right)_{ \b} { S }^\m_m \} = \epsilon_{ \a \b} \CL_2 \, ,
\label{AB:exact} \ee
where $f_{ \b}^{+} $ includes the sum of the terms which appeared in the $Q$-exact parts of the A- and B-models,
\be f^{+}_{ \b} \equiv 2 \omega_{ AB} \left( ( \chi_\m^A D^{ + \m  } q^B_{ \b } - i H^A_{ \b } \lambda^B )   - \tilde{ \zeta}^A_{ \b } ( \Ds^{+} \tilde{ q}^B + 2 \tilde{H}^B - \tilde{h}^B ) \right)  \ .
\ee
The other two terms in \eqref{AB:exact} are novel in the AB-model.

\subsection{Gauge fixing}

\paragraph{Gauge fixing} Let us consider the BRST quantization of the AB-model. The BRST quantization of A-model has been considered in \cite{ks}. In there, it has been noticed that imposition of a non-trivial variation of the ghost  $c^m$ with respect to the holomorphic super-charge, $Q_+$ in our notation, can result in a nilpotent scalar charge. We will see that the same prescription also  works for the AB-model. Reduction to the A- or B-models can be done trivially.

For the quantization,  we introduce fermionic ghost and anti-ghost,  $c^m, \bar{c}^m $, and bosonic ghost $B^m$. Recall the standard Fadeev-Popov BRST variations for all fields,
\begin{align}
& [  \hat{Q}, q^A_{ \alpha } ] =  - ( t_m)^A_{ \ B } q^B_{ \a } c^m, \quad \{ \hat{Q}, \lambda^A \} = ( t_m)^A_{ \ B } \lambda^B c^m, \quad \{ \hat{ Q} , \chi_{ \mu}^A  \} = ( t_m)^A_{ \ B } \chi_{ \mu}^B c^m , \nn \\
& [ \hat{Q} , \tilde{q}^A ] =  - ( \tilde{t}_m )^A_{ \ B }  \tilde{q}^B c^m \ ,    \quad  \{ \hat{Q} , \tilde{ \zeta}^A_{ \a} \} = ( \tilde{t}_m)^A_{ \ B } \tilde{ \zeta}^B_{ \a }  c^m , \quad [ \hat{Q} , \tilde{h}^A ] = - ( \tilde{t}_m)^A_{ \ B } \tilde{h}^B c^m ,    \nn \\
& [ \hat{Q} , A_\m^m ] =  D_\m c^m,  \quad  \{ \hat{Q} , c^m \} = - \half f^m_{ \ n p } c^n c^p , \quad \{ \hat{Q} , \bar{c}^m \} = B^m, \quad [ \hat{Q} , B^m ] = 0 .
\end{align}
$\hat{Q}^2$ is nilpotent as in the standard BRST quantization.
Now, following \cite{ks}, we impose the following super-symmetry variations of ghost fields, where $ Q \equiv Q_{+} $,
\be \{ Q ,  c^m \} = - \frac{i}{2} \mu^{m}_{++}, \quad \{ Q , \bar{c}^m \} = 0 , \quad [ Q, B^m ] = 0 , \label{q:ghost}
\ee
then $\hat{Q}$ no longer anti-commutes with $ Q$.   The non-vanishing variation in \eqref{q:ghost} is chosen in the way that the non-anti commuting part of $Q$ and $ \hat{Q}$ can cancel the remnants  of $Q^2$ in  $ ( Q + \hat{Q})^2 $ : For $\CQ \equiv Q + \hat{Q}$ , the following holds for all fields
\be [ \{ \CQ, \CQ \} , \,\cdot\, ] = 0 . \ee

 Consider a gauge fixing function $f^m(A)$, for instance, $f^m(A) = \p^\m A_\m^m $ for the Lorentz gauge. The gauge-fixing  term appears in a $ \CQ$-exact form,
 \be
 \CL_{\rm g.f.} = \{ Q,  \bar{ c}_m f^m (A ) \} \  ,   \ee
thus the total Lagrangian $ \bar{ \CL} = \CL_1 + \CL_2 + \CL_{\rm g.f.}$ is $\CQ$-closed.

\section{Relation to other topological theories \label{last}}

We have carried out the (A- and B-) topological twisting 
of general $\CN=4$ supersymmetric Chern-Simons gauged sigma models
of \cite{gw,n4}. The next task is to evaluate the 
partition function and correlation functions of the quantum theory, 
which should provide (hopefully new) topological invariants of three manifolds. 

The computation involves roughly three steps; see, for instance, \cite{rev}. 
First, one should identify observables, which are elements of $Q$-cohomology. 
Second, the path integral often localizes onto the moduli space of $Q$-invariant configurations (``instantons"). Third, perturbation theory 
is used to compute the partition function unless a more powerful tool is available. 

In this section, we point out some interesting connections 
between our new theories and previously well studied theories in the literature. 
We believe that our findings will 
serve as a useful guide in taking each of the three step of computation, 
most of which we leave for a future work.  

For the readers' convenience, we begin with a brief review 
of well-known topological theories in three dimensions 
and how they are related to each other.  
Next, we review a useful fact about (physical) Chern-Simons-matter theories 
which inspired our main observation. It is the generalized Higgs mechanism 
of Mukhi-Papageorgakis (MP) \cite{mukhi} which, to some extent, transforms 
Chern-Simons theories to Yang-Mills theories. We will see that the MP map 
is compatible with the topological twisting and 
that some main results of the previous sections 
have dual interpretations on the Yang-Mills side.  

In the last subsection, we will combine everything 
and try to obtain some clues as to what topological 
invariants our new theories may compute. 
In particular, we will argue that the A-model 
should capture the Casson invariant 
and speculate on the role of a complexified gauge group 
in the B-model.  

\subsection{Review of old results \label{old-tft}} 

There are largely three well-known three dimensional TFTs in the literature;
pure Chern-Simons theory \cite{purecs},
A-twisted $\CN=4$ super Yang-Mills theory \cite{cass1,cass2},
and A-twisted ungauged $\CN=4$ sigma model of Rozansky-Witten \cite{rw}.
Let us review some features of
these theories relevant for our discussion below.

The A-twisted Yang-Mills theory is known to compute Casson invariant which is,  roughly speaking, a signed sum over flat connections ($F_{\m\n}=0$).
An important fact for our discussion is that Casson invariant admits
an alternative field theory description \cite{cass1}. It is a sort of super-BF theory whose Lagrangian is, in a differential form notation,
\be
\label{sBF}
\CL = \tr \left(B\wedge F_A + \chi \wedge d_A \psi \right) \,.
\ee
Here, $A$ is the gauge connection and $F_A$ the curvature. $B$ is a bosonic one-form and $\chi$, $\psi$ fermionic one-forms, all of which are Lie-algebra-valued. The Yang-Mills and BF descriptions look very different at first sight, but it was shown \cite{cass2} that the Yang-Mills theory can be deformed
in a topologically invariant way to the BF theory.

Rozansky-Witten theory \cite{rw} is not a gauge theory. Nevertheless, it is intimately related to both pure Chern-Simons and Yang-Mills theories.

First, it computes the $SU(2)$ Casson invariant if one chooses the target space
to be the Atiyah-Hitchin space. The physical explanation is that
the low energy limit of the (physical) Yang-Mills theory is a sigma model
on the moduli space of vacua. Taking account of loop and instanton effects,
the quantum moduli space of vacua of the $SU(2)$ super Yang-Mills theory
has been shown to be the Atiyah-Hitchin space \cite{sw3}. 
\footnote{
See \cite{gates3} for a discussion of the super-BF theory 
in a manifest $\CN=4$ language.}

Second, the relation to pure Chern-Simons theory can be seen
by comparing the perturbation theory of Rozansky-Witten theory
to that of Chern-Simons theory. Topological invariance of amplitudes
in perturbation theory is verified by the so-called IHX relation
\cite{natan}. The essence of the IHX relation for Chern-Simons theory
is the Jacobi identity of Lie algebra, while a similar role is played by the Bianchi identity for the Riemann curvature in Rozansky-Witten theory. With this formal similarity in mind, a detailed comparison of Feynman diagrams in the two theories
reveals certain relations between Chern-Simons and Casson invariants.

Relatively less known, but equally important for our discussion, 
is the B-twisted $\CN=4$ super Yang-Mills theory \cite{bth}. 
We will review some aspects of this theory in the next subsection. 

\subsection{MP map : Chern-Simons vs Yang-Mills \label{mp}}

We now present the MP map \cite{mukhi} customized to the linear model with $U(N)\times U(N)$ gauge symmetry and bi-fundamental matter fields
\begin{eqnarray}
  Z_\a , \Psi^\da\  : \ \ ({\bf N}, \bar {\bf N})\ , \qquad
  \bar Z^\a, \Psi_\da : \ \ (\bar {\bf N}, {\bf N})\ ,
\end{eqnarray}
where $\a=1,2$ and $\da=1,2$ denote the $R$-symmetry indices as before.

In the $\CN=2$ super-field notation, the scalar fields split into two charged chiral super-fields $Z_\a =( A,B^\dagger )$.  The $\CN=4$ vector multiplets decomposes into vector super-fields $V_B$ ($V_C$) combined with
and two neutral chiral super-fields $\Phi_B, \Phi_C$.

For simplicity, let us discuss the abelian (Maxwell) case first.
The $\CN=4$ Lagrangian of the Gaiotto-Witten model in the $\CN=2$ notation 
is
\begin{eqnarray}\label{Lagrangian2}
  \CL = \int d^4\th \left( A^\dagger e^{-2V_B} A
  + B e^{2V_B} B^\dagger + 2 V_B \Sigma_C
  - \left[\int d^2\th \  \Phi_B \Phi_C
  - \sqrt2 B \Phi_B A  + \text{ c.c.} \right] \right)\ ,
\end{eqnarray}
where $\S_{B,C}$ are the field strength super-fields for $V_{B,C}$, defined as
\begin{eqnarray}
  \Sigma = -\frac{i}{2} \e^{\a\b} D_\alpha
  \bar D_\beta V \,.
\end{eqnarray}
They satisfy the defining relations of a linear superfield
\begin{eqnarray}\label{lsuperfield} 
  D^2 \Sigma = \bar D^2 \Sigma =0 \,.
\end{eqnarray}
%

The equations of motion for auxiliary superfields $V_B$ and $\Phi_B$ are
\begin{eqnarray}\label{a}
  e^{-2V_B} A A^\dagger  - e^{2V_B} B^\dagger B
  - \Sigma_C=0 , \qquad
  \Phi_C = \sqrt2 A B \ .
\end{eqnarray}
Integrating out the auxiliary fields, we recover
the Lagrangian in the main text with Yukawa-like term
and sextic bosonic potential.
Instead, we may choose to eliminate the matter multiplets by solving for $e^{V_B}$,
\begin{eqnarray}\label{b}
  e^{-2V_B} = \frac{
   \Sigma_C \pm \sqrt{(\Sigma_C)^2+4|B A|^2}}
  {2A^\dagger A}, \qquad
  V_B = -\frac12 \log{\left(
  \frac{\S_C \pm \sqrt{\S_C^2 + 2|\Phi_C|^2}}
  {2 A^\dagger A}
  \right)}\ ,
\end{eqnarray}
and inserting it back into (\ref{Lagrangian2}). 
Taking account of the chirality of $A,A^\dagger$ and the 
properties of linear superfields $\Sigma_C$ (\ref{lsuperfield}), one can 
obtain the dual vector description of the model, 
\begin{eqnarray}\label{Lagrangian4}
  \CL_\text{dual} = \int d^4\th \left(
  \sqrt{\Sigma_C^2 +  2|\Phi_C|^2} - \Sigma_C \log{\left[
  \Sigma_C + \sqrt{\Sigma_C^2 + 2|\Phi_C|^2}
  \right]} \right) \ ,
\end{eqnarray}
whose bosonic parts take the form
\begin{eqnarray}\label{Lagangiandual2}
  \CL= -\frac{1}{2\sqrt{\phi^2+ 2|q|^2}} \left( \frac14 F_{\mu\nu}^2 +
  \frac12 (\partial_\mu \phi)^2 + |\partial_\mu q|^2 \right) - \frac{1}{2\pi}
  \e^{\mu\nu\rho}F_{\mu\nu} \omega_\rho\ .
\end{eqnarray}
Here, $\phi$ and $q$ are the lowest components of $\Sigma_C$ and $\Phi_C$, 
and $\omega_\mu$ is the pull-back of the Dirac vector potential in the target
space $(\phi,q,\bar q)\in \mathbb{R}^3$:
\begin{eqnarray}
  \omega_\mu = \frac12 (1-\cos{\th}) \partial_\mu \varphi. \qquad \big(\phi=r\cos{\th}, \
  q=\frac{1}{\sqrt2} r\sin{\th}e^{i\varphi}\big)\  .
\end{eqnarray}
Restoring the dependence of Chern-Simons level $k$, one can conclude that
the Coulomb branch of (\ref{Lagangiandual2}) is the orbifold
$\mathbb{C}^2/\mathbb{Z}_k$, the same as the moduli space of the vacua 
of the $U(1)\times U(1)$ Gaiotto-Witten model.


%
To complete the MP map, we give a vev to $\S_C$ by setting $\S_C = v^2 + \d \S_C$. The leading behavior of (\ref{Lagrangian4}) with respect to $1/v^2$
is
\begin{eqnarray}
  \CL &\simeq&
  -\frac{1}{2v^2} (\d \S_C)^2 + \frac{1}{2v^2} (\Phi^\dagger \Phi)
  + \CO(\frac{1}{v^2})
  \nn \\
  &\goto&  -\frac{1}{g^2} \left(
  \frac14 F_{\mu\nu}^2 + \frac12 \sum_{i=1}^3 (\partial_\mu \phi_C^i)^2
  - i \sum_{a=1}^2 \bar \l^a \g^\mu \partial_\mu \l_a \right), 
\label{max4}
\end{eqnarray}
with $1/g^2 = 1/2v^2$. It is precisely
the $\CN=4$ supersymmetric Maxwell theory.

%

The same idea can be applied to the $U(N)\times U(N)$ theories,
but the computations are more involved. We jump
directly to the final identification between the hyper-multiplets
of the Chern-Simons side and the vector-multiplets of the Yang-Mills side,
\begin{eqnarray}\label{mpMap}
  \l_{\a\da} &=& \frac12 \Big[  \big(
  Z_\a \bar \Psi_\da + \e_{\a\b}\e_{\da\db} \Psi^\db \bar Z^\b
  \big)  + (\text{un-bar} \leftrightarrow \text{bar}) \Big]\ ,
 \nn \\
  \vec \Phi &=& \frac12 \Big[
  Z_\a \vec \t_\b^{\ \a} \bar{ Z^\b} + ( \text{un-bar} \leftrightarrow \text{bar}
  ) \Big]\ ,
\end{eqnarray}
up to the leading order of $1/v^2$ expansion. 
The MP map can be understood as a non-abelian version of the vector-scalar 
duality in three dimensions.  

%
%

\paragraph{Compatibility I. A-twisting}
To show that the MP map is compatible with the A-twisting,
we begin by recalling the supersymmetry transformation rules
of super Yang-Mills theory,
\begin{eqnarray}
  && \hspace*{2cm}\d \vec \Phi = \eta^{\a\da} \vec \t_\a^{\ \b} \l_{\b \da}\ ,
  \qquad \d A_\mu = i \eta^{\a\da} \g_\mu \l_{\a\da}\ ,
  \nn \\
  &&  \d \l_{\a \da} =
  \frac12 F_{\mu\nu} \rho^{\mu \nu} \eta_{\a \da}
  + i \vec \t_\a^{\ \b} \rho^\mu \eta_{\b \da} \cdot  D_\mu \vec \Phi
  + \frac12 \e_{mnp} \big[ \Phi^m, \Phi^n \big]
  \big(\tau^p \big)_\a^{\ \b}\eta_{\b \da}
\end{eqnarray}
The twisting is done by the following substitutions,
\begin{eqnarray}
  \big( \eta^s \big)_\a^\da = -\frac12 \e^{s\da} \eta_\a\ ,
  \qquad
  \big(\l_s\big)_{\a\da} = \e_{s\da} \l_\a +
  \big( \g^\mu \e \big)_{s\da} \chi_{\mu \a}\ ,
\end{eqnarray}
which yields the transformation rules for the topological theory,
\begin{eqnarray}
  && \hspace*{3cm}\d \vec \Phi = \eta^\a \vec \t_\a^{\ \b} \l_\b\ ,
  \qquad \d A_\mu = - i \eta^\a \chi_{\mu\a}\ , \nn \\
  && \d \l_\a = \frac14 \e_{mnp} \big[ \Phi^m ,\Phi^n \big]
  \big( \t^p \big)_\a^{\ \b} \eta_\b \ , \qquad
  \d \chi_{\mu\a} = \frac{i}{2}\Big[ \frac12 \e_{\mu\nu\rho}
  F^{\nu\rho} \d_\a^{\ \b}
  +  D_\mu \vec\Phi \cdot \vec \t_\a^{\ \b}  \Big] \eta_\b\ .
\end{eqnarray}
These results should be compared
to the transformation rules of the A-twisted Chern-Simons theory
inherited through the MP map. For scalar fields, we find
 \begin{eqnarray}
   \d \vec \Phi = i \vec \t_\a^{\ \b} \eta^\a \frac{1}{2}\big(
   Z_\b \bar \l + \e_{\b\g} \l \bar Z^\g \big) + ( \text{un-bar} \leftrightarrow \text{bar}
  ) =
   i \vec \t_\a^{\ \b} \eta^\a \l_\b\ .
 \end{eqnarray}
The field $\chi_{\m\a}$ is written in terms of matter fields
of Chern-Simons theory as
 \begin{eqnarray}
   \chi_{\m\a} = \frac{1}{2} \big( Z_\a \bar \chi_\m
   + \e_{\a\b} \chi_\m \bar Z^\b \big) + ( \text{un-bar} \leftrightarrow \text{bar}
  ) \ .
 \end{eqnarray}
 Applying the BRST transformation rule of A-model,
 one can obtain
 \begin{eqnarray}
   \d \chi_{\m\a} &=&
   \eta_\a \frac{1}{2}\Big[ \big(i (
   \l \bar \chi_\m + \chi_\m \bar \l ) - \frac14 \e^{\b\g}
   (Z_\b \cdot D_\m \bar Z_\g + D_\m Z_\b \cdot \bar Z_\g) \big) \nonumber \\
   && \hspace*{1cm}
   + \eta^\b \big(\frac{1}{2} D_\m \Phi_{(\a} \Phi_{\b)}\big)\Big]  +
   (\text{un-bar} \leftrightarrow \text{bar}) \nonumber \\
   &=& -\frac12 \left[ \frac12 \e_{\m\n\l} F^{\n\l} \e_\a^{\ \b}
   + D_\m \vec \Phi \cdot \vec \t_\a^{ \b} \right] \eta_\b\ ,
 \end{eqnarray}
 where we used the Gauss law in the last step. One can easily
show that the rests of transformation rules of SYM
are also uncovered by the same manner.

\paragraph{Instanton of the A-model}

The BRST transformation implies that the supersymmetric configurations
should satisfy
\begin{eqnarray}
  D_\mu Z_\a = D_\mu \bar Z^\a = 0 \ ,\qquad
  H^\a = \bar H_\a = 0\ ,
\end{eqnarray}
where
\begin{eqnarray}
  H^\a =
  Z_1 \bar Z^\a Z_2 - Z_2 \bar Z^\a Z_1 \ ,
  \qquad
  \tilde H_\a =
  \bar Z^1 Z_\a \bar Z^2 - \bar Z^2 Z_\a \bar Z^1 \ .
\end{eqnarray}
The Gauss law becomes
\begin{eqnarray}
 \ast F =  D Z_\a \bar Z^\a - D Z_\a \bar Z^\a\ .
\end{eqnarray}
As first shown in \cite{m2-inst}, through the MP map,
these equations can be transformed
into the familiar instanton equations of the Yang-Mills theory.
For instance, the supersymmetric configurations of A-model
implies those of A-model of super Yang-Mills:
\begin{eqnarray}
  D_\mu Z_\a = D_\mu \bar Z^\a = 0 \ \&
  \ \text{Gauss law} \ \to \
  F_{\mu\nu} = 0 \ , \ \
  D_\mu \vec \Phi = 0 \ .
\end{eqnarray}
The F-term equation also translates into
\begin{eqnarray}
  H^\a = \tilde H_\a = 0 \ \to \
  \big[ \Phi^m , \Phi^n \big] = 0 \ .
\end{eqnarray}
Furthermore, a half-BPS instanton can be written down concretely.
For a solution to keep $\eta_2$ unbroken, we have to require
\begin{eqnarray}
  D_\mu Z_1 = D_\mu \bar Z^2 = 0 , \qquad
  Z_2= \bar Z^1 = 0 \ .
\end{eqnarray}
They imply that F-term conditions are automatically
satisfied and
\begin{eqnarray}
  \text{Gauss law} \ \to \ \ast F = D \Phi^3 \ , \
  \Phi^1 = \Phi^2 = 0 \   \qquad
  \big( \Phi^3 = \frac12 \big(Z_1 \bar Z^1 - Z_2 \bar Z^2 + \cdots\big)  \Big) \ ,
\end{eqnarray}
These are nothing but the equations the half-BPS instanton of
A-model SYM should satisfy.

\paragraph{Compatibility II. B-twisting}

The $\CN=4$ super Yang-Mills theory in three dimensions
also allow for the B-twisting \cite{bth}. Concretely,
\begin{eqnarray}
  \big( \eta^s \big)^{\a\da} = \frac12 \e^{s\a} \eta^\da\ ,
  \qquad
  \big(\l_s\big)_{\a\da} = \e_{s\a} \l_\da +
  \big( \g^\m \e \big)_{s\a} \chi_{\m \da}\ .
\end{eqnarray}
After the twisting, one can obtain the following BRST
transformation rules:
\begin{eqnarray}
  &&\d \vec \Phi_\m = \eta^\da \chi_{\m\da}\ ,
  \qquad
  \d \l_\da = \frac{i}{2} D^\m \Phi_\m \eta_{\da} \ ,
  \qquad
  \d A_\m = i \eta^{\da} \chi_{\m\da}\ ,
  \nn \\
  &&\d \chi_{\m\da} = - \frac{\e^{\m\n\l}}{2} \Big(
  \frac{i}{2}F_{\n\l} + D_\n \Phi_\l - \frac12 \big[ \Phi_\n,
  \Phi_\l \big] \Big) \eta_{\da}
\end{eqnarray}
Applying the BRST transformation rules of B-model (\ref{Btransf})
together with the MP map (\ref{mpMap})
gives us again the above transformation rules as expected.

\paragraph{Instanton of the B-model}

For the B-twisted super Yang-Mills, the supersymmetric configurations
satisfy
\begin{eqnarray}\label{BPS2}
  \frac i2 F_{\m\n} + D_{[\m} \Phi_{\n]} - \frac12
  \big[ \Phi_\m, \Phi_\n \big] = 0\ , \qquad  D^\m \Phi_\m = 0\ .
\end{eqnarray}
Once we define a twisted {\em complex} gauge field $A_\mu^+$ by
\begin{eqnarray}
  A^+_\mu = A_\mu - i \Phi_\mu \ ,
\end{eqnarray}
the first equation of (\ref{BPS2}) can be rewritten as
\begin{eqnarray}
  F^+_{\m\n} = 0 \ ,
\end{eqnarray}
Clearly,
this modification of the gauge field exactly
parallels what we found in section 3. In particular, 
as pointed out in \cite{bth}, the B-twisted super Yang-Mills carries the Wilson line as a topological observable,
\begin{eqnarray}
  W(C) = \tr\left( \CP \exp \left[\oint_C dx^\m (A_\m - i\Phi_\m) \right] \right)\ , \qquad
\d(A_\m - i \Phi_\m) = 0 \ .
\end{eqnarray}

\subsection{Discussions}

\paragraph{A-model}

In subsection \ref{old-tft}, we mentioned that 
the A-twisted super Yang-Mills theory computes Casson invariant.  
Let us restrict our attention to the simplest gauge group $SU(2)$. 
Via the MP map, it corresponds to the $SU(2)\times SU(2)$ linear model.
Therefore, we suspect that the latter should also capture the $SU(2)$ Casson invariant. Let us give some further heuristic arguments supporting this intriguing possibility.

The first step in taking the MP map is to give a constant non-zero vev
for the scalar field in the Chern-Simons sigma model:
\be
\label{mp-vev}
Z_\a = v_\a \identity_{2\times 2}\,,
\ee
which leads to (\ref{max4}) through (\ref{a}). 
We use the notation in which the gauge group acts on $Z$ by
$Z \goto U \Phi \widetilde{U}^{-1}$.
The vev (\ref{mp-vev}) makes sense
even when the three-manifold and the gauge bundle have
non-trivial topology as long as the two $SU(2)$ bundles
share the same topology; the choice of vev (\ref{mp-vev}) further
assumes the same trivialization for the two bundles.
The precise value of $v_\a$ is not important
because the $R$-symmetry and scale invariance relate any non-zero value of
the vev. The vev (\ref{mp-vev}) breaks the gauge group
into the diagonal subgroup, under which the matter fields transform
in the adjoint representation.
Now, note that the $Q$-closed part of the Lagrangian of the A-model
(\ref{A_CL1}) reduces precisely to the super-BF theory
(\ref{sBF})!
\footnote{
There is another parallel between the two theories.
The super-BF theory (\ref{sBF}) was originally introduced
as a Chern-Simons theory whose gauge group is a super-group \cite{cass1}.
The same interpretation was given to the A-model
in \cite{ks}.
}

Assuming that the contribution from the conformal (zero-vev) point 
does not spoil everything,
our arguments would imply that the A-model offers a new way to understand
the equivalence \cite{cass2} between super-Yang-Mills and
super-BF theories, both of which compute Casson invariant.

The MP map also gives a hint on how to carry over the relation between
Yang-Mills and Rozansky-Witten theories to the Chern-Simons sigma model
context. In showing that
the moduli space of the vacua of the (physical) $SU(2)$ Yang-Mills theory
is the Atiyah-Hitchin space, the correction to the moduli space metric
due to instantons is crucial.
As we showed above, the Chern-Simons sigma model shares the same instantons as the Yang-Mills theory.
The argument based on fermion zero mode counting \cite{sw3}
seems to work equally well in the Chern-Simons setup,
so we find it plausible that
in the ``off-diagonal'' part of the moduli space of vacua
again becomes the Atiyah-Hitchin space.


Perturbation theory may illuminate different aspects of the A-model.  
A crucial step in the perturbative analysis is to verify topological invariance by the so-called IHX relation \cite{natan}. 
As emphasized in \cite{ks}, the A-model is a combination of pure Chern-Simons and Rozansky-Witten theories. 
It would be very interesting to figure out how the gauge fields and matter fields 
conspire to give a new example of the IHX relation. 


\paragraph{B-model}

In the B-model, the relation to pure Chern-Simons could be more direct,
since the topological part of the B-model Lagrangian is already
a pure Chern-Simons action, albeit with a modified gauge field; see
(\ref{lag_B}) and (\ref{CA}). The shift of the gauge field
is purely imaginary and leads to a complex gauge field.
Of course, this does not imply complexification of the underlying gauge symmetry. 
Perhaps surprisingly, however, complexified gauge symmetry does play a role in 
some context. 

Marcus \cite{marc} studied a twisted $D=4$, $\CN=4$ super Yang-Mills theory and showed that the instanton equation is the flat connection 
condition for the complex gauge field: $F_{\m\n}^+ = 0$. Moreover, 
he showed that complexified gauge group somehow plays an important role  
in understanding the moduli space of flat connections. 
An interpretation of this observation was given by Baulieu \cite{baul} 
who re-interepreted the $\CN=4$ Yang-Mills theory as 
a complexification of the $\CN=2$ theory. 

Whether a similar story holds for our B-model and its Yang-Mills cousin 
is an open question. 
If so, the B-model may even be related to recent developments (see, {\it e.g.}, \cite{guk} and references therein) in pure 
Chern-Simons theories with complex gauge groups.

\paragraph{Quantization} 
Finally, it may be helpful to study Batalin-Vilkovisky (BV) quantization
of the new theories as the BV quantization facilitates comparison
between different gauges which illuminates different aspects of the same theory. The BV quantization of Rozansky-Witten theory has been done recently \cite{qz}
following the AKSZ prescription \cite{aksz}.

\vskip 1.5cm

\section*{Acknowledgments}
We are grateful to Young-Hoon Kiem, Seok Kim, Jaemo Park, Jae-Suk Park, Jeong-Hyuck Park, Ki-Myeong Lee, Ho-Ung Yee, Satoshi Yamaguchi and Piljin Yi for discussions and comments on the manuscript. We also thank Kazuo Hosomichi for collaboration
at an early stage of this work, and Kentaro Hori and Nikita Nekrasov
for discussions in summer 2008.
The work of EK is supported in part by KRF-2005-084-C00003, EU FP6 Marie Curie Research
\& Training Networks MRTN-CT-2004-512194 and HPRN-CT-2006-035863 through MOST/KICOS.
The work of
Sm.L. is supported in part by the KRF/KOSEF Grants KRF-2007-331-C00073 and KOSEF-2009-0072755.


\vskip 1.5cm

\centerline{\large \bf Appendix}

\appendix


\section{Some details of computations \label{app:detail}}

\paragraph{Notations on spinors}
Spinor indices run $\alpha=+,-$. Indices are raised or lowered
by real antisymmetric matrices $\epsilon_{\alpha\beta}$ and
$\epsilon^{\alpha\beta}$ satisfying
$\epsilon_{\alpha\beta}\epsilon^{\beta\gamma}=\delta_\alpha^{~\gamma}$.
\[
 \psi_\alpha = \epsilon_{\alpha\beta}\psi^\beta,~~~~~
 \psi^\alpha = \epsilon^{\alpha\beta}\psi_\beta.
\]
Space-time metric has signature $(-++)$.
The $\gamma$-matrices $(\gamma^\mu)_\alpha^{~\beta}$ satisfy the relations
\[
 \gamma^\mu\gamma^\nu+\gamma^\nu\gamma^\mu=2\eta^{\mu\nu},~~~~~
 \gamma^{[\mu}\gamma^\nu\gamma^{\rho]}=\varepsilon^{\mu\nu\rho}.~~~~~
 (\varepsilon^{012}=1)
\]
The matrices $(\epsilon\gamma^\mu)^{\alpha\beta}$ are real symmetric.
Vectors such as $x^\mu$ and $\partial_\mu$ are expressed as bi-spinors
\[
 x^{\alpha\beta}= x^\mu(\epsilon\gamma_\mu)^{\alpha\beta},~~~~
 \partial_{\alpha\beta}=-(\gamma^\mu\epsilon)_{\alpha\beta}\partial_\mu.
\]
Spinor indices in the standard position will be omitted.
\[
 \psi\theta\equiv\psi^\alpha\theta_\alpha,~~~~
 \theta^2=\theta^\alpha\theta_\alpha,~~~~
 \psi\gamma^\mu\theta=\psi^\alpha(\gamma^\mu)_\alpha^{~\beta}\theta_\beta,~~~~
 {\rm etc}.
\]

\paragraph{Nilpotency of the A-model} Let us consider some details in \eqref{nil1}.
\be
 [ \{ Q_{ \a } , Q_{ \b }  \} , q^A_{ \g } ] &=& - i ( \epsilon_{ \b \g } H^A_{ \a } + \epsilon_{ \a \g } H^A_{ \b } )  \nn \\
& = & - i ( t^m)^A_{ \ B } ( \mu_m)_{ \a \b } q^B_{ \g} .
\ee
In the second line, we used
\be \epsilon_{ \a \b } q^A_{ \g} + \epsilon_{ \b \g } q^A_{ \a} + \e_{ \g \a } q^A_{ \b} = 0 . \ee
In checking the nilpotency on $ \lambda^A$, it is useful to note that
\be [ Q_{ \a } , H^A_{ \b } ] = - \half ( t^m)^A_{ \ B } ( \mu_m)_{ \a \b } \lambda^B - \half \e_{ \a \b} (t^m)^A_{ \ C } (t_m)_{ BD } q^C_{ \g } q^{ D \g } \lambda^B .
\ee
In checking nilpotency of $ \tilde{h}^A$ in the AB-model \eqref{nil:AB}, the following relation can be useful
\be  ( \tilde{t}^m)^A_{ \ B } ( t_m)_{ CD} q^C_{ \b} H^D_{ \a } \tilde{q}^B + ( \a \leftrightarrow \b ) = - \left( \frac{1}{4} f_{ mnp} ( \tilde{t}^m)^A_{ \ B } ( \mu^p )^{ \g}_{ \ \a } ( \mu^n )_{ \g \b } \tilde{q}^B + ( \a \leftrightarrow \b ) \right)  .
\ee

\paragraph{Proof of an identity} Let us denote the left hand side of \eqref{pf} by $ \CB$. Then
\be \CB &=& \half  S^m_\m ( \tilde{t}_m)_{ AB} \tilde{q}^A \{ \gamma^\m , \gamma^\n \} D_\n \tilde{q}^B \nn  \\
& = & \half ( \tilde{t}^m)_{ AB} ( \tilde{t}_m)_{ CD} \left( ( \tilde{q}^C \Ds \tilde{q}^B ) ( \tilde{q}^A \tilde{q}^D ) + ( \tilde{q}^A \gamma^\m \tilde{q}^D ) ( \tilde{q}^C D_\m \tilde{q}^B )  \right)  \label{b.1}
 \, .  \ee
where we used  the following Fierz identity,
\be ( \gamma^\m )_{ \a }^{ \ \b} ( \gamma_\m )_{ \s}^{ \ \d} = 2 { \d }_{ \s}^{ \b} { \d}_{ \a}^{ \d} - { \d}_{ \a}^{ \b} { \d}_{ \s}^{ \d}  \ .
\ee
The second term of \eqref{b.1} becomes
\be \half ( \tilde{t}^m)_{ AB} ( \tilde{t}_m)_{ CD}  ( \tilde{q}^A \gamma^\m  \tilde{q}^D ) ( \tilde{q}^C D_\m \tilde{q}^B )
& = & - \CB - \CB +\half  ( \tilde{t}^m)_{ AB} ( \tilde{t}_m)_{ CD} ( \tilde{q}^C \Ds \tilde{q}^B ) ( \tilde{q}^A \tilde{q}^D )
 , \nn \ee
 while
 \be ( \tilde{t}^m)_{ AB} ( \tilde{t}_m)_{ CD}  ( \tilde{q}^C \Ds \tilde{q}^B ) ( \tilde{q}^A \tilde{q}^D ) = 3  \omega_{ AB} \tilde{H}^A \Ds \tilde{q}^B  \ .  \ee
Thus \eqref{pf} is proved.

\end{document}